\documentclass[]{bmcart}

\usepackage[utf8]{inputenc} %unicode support

\usepackage{graphicx} %to remove for submission
\usepackage{hyperref}

\usepackage{soul} %strikethrough text

\startlocaldefs
\endlocaldefs

\begin{document}

%%% Start of article front matter
\begin{frontmatter}

\begin{fmbox}
\dochead{Research}


\title{A Large Scale Study of Reader Interactions with Images on Wikipedia}

%%%%%%%%%%%%%%%%%%%%%%%%%%%%%%%%%%%%%%%%%%%%%%
%%                                          %%
%% Enter the authors here                   %%
%%                                          %%
%% Specify information, if available,       %%
%% in the form:                             %%
%%   <key>={<id1>,<id2>}                    %%
%%   <key>=                                 %%
%% Comment or delete the keys which are     %%
%% not used. Repeat \author command as much %%
%% as required.                             %%
%%                                          %%
%%%%%%%%%%%%%%%%%%%%%%%%%%%%%%%%%%%%%%%%%%%%%%

\author[
   addressref={aff1},                   % id's of addresses, e.g. {aff1,aff2}
   corref={aff1},                       % id of corresponding address, if any
   %noteref={n1},                        % id's of article notes, if any
   email={daniele.rama@unito.it}   % email address
]{\inits{DR}\fnm{Daniele} \snm{Rama}}
\author[
   addressref={aff2},
   email={tiziano.piccardi@epfl.ch}
]{\inits{TP}\fnm{Tiziano} \snm{Piccardi}}
\author[
   addressref={aff3},
   email={miriam@wikimedia.org}
]{\inits{MR}\fnm{Miriam} \snm{Redi}}
\author[
   addressref={aff1,aff4},
   email={rossano.schifanella@unito.it}
]{\inits{RS}\fnm{Rossano} \snm{Schifanella}}

%%%%%%%%%%%%%%%%%%%%%%%%%%%%%%%%%%%%%%%%%%%%%%
%%                                          %%
%% Enter the authors' addresses here        %%
%%                                          %%
%% Repeat \address commands as much as      %%
%% required.                                %%
%%                                          %%
%%%%%%%%%%%%%%%%%%%%%%%%%%%%%%%%%%%%%%%%%%%%%%

\address[id=aff1]{%                           % unique id
  \orgname{University of Turin},
  \city{Turin},                              
  \cny{Italy}                                    
}
\address[id=aff2]{%
  \orgname{École polytechnique fédérale de Lausanne (EPFL)},
  \city{Lausanne},
  \cny{Switzerland}
}
\address[id=aff3]{%
  \orgname{Wikimedia Foundation},
  \city{San Francisco, CA},
  \cny{US}
}
\address[id=aff4]{%
	\orgname{ISI Foundation},
	\city{Turin},
	\cny{IT}
}

%%%%%%%%%%%%%%%%%%%%%%%%%%%%%%%%%%%%%%%%%%%%%%
%%                                          %%
%% Enter short notes here                   %%
%%                                          %%
%% Short notes will be after addresses      %%
%% on first page.                           %%
%%                                          %%
%%%%%%%%%%%%%%%%%%%%%%%%%%%%%%%%%%%%%%%%%%%%%%

\begin{artnotes}
%\note{Sample of title note}     % note to the article
%\note[id=n1]{Equal contributor} % note, connected to author
\end{artnotes}

\end{fmbox}% comment this for two column layout

%%%%%%%%%%%%%%%%%%%%%%%%%%%%%%%%%%%%%%%%%%%%%%
%%                                          %%
%% The Abstract begins here                 %%
%%                                          %%
%% Please refer to the Instructions for     %%
%% authors on http://www.biomedcentral.com  %%
%% and include the section headings         %%
%% accordingly for your article type.       %%
%%                                          %%
%%%%%%%%%%%%%%%%%%%%%%%%%%%%%%%%%%%%%%%%%%%%%%

\begin{abstractbox}

\begin{abstract}
Wikipedia is the largest source of free encyclopedic knowledge and one of the most visited sites on the Web. To increase reader understanding of the article, Wikipedia editors add images within the text of the article's body. However, despite their widespread usage on web platforms and the huge volume of visual content on Wikipedia, little is known about the importance of images in the context of free knowledge environments. To bridge this gap, we collect data about English Wikipedia reader interactions with images during one month and perform the first large-scale analysis of how interactions with images happen on Wikipedia. 
First, we quantify the overall engagement with images, finding that one in $29$ pageviews results in a click on at least one image, one order of magnitude higher than interactions with other types of article content. Second, we study what factors associate with image engagement and observe that clicks on images occur more often in shorter articles and articles about visual arts or transports and biographies of less well-known people. Third, we look at interactions with Wikipedia article previews and find that images help support reader information need when navigating through the site, especially for more popular pages. The findings in this study deepen our understanding of the role of images for free knowledge and provide a guide for Wikipedia editors and web user communities to enrich the world's largest source of encyclopedic knowledge.
\end{abstract}

%%%%%%%%%%%%%%%%%%%%%%%%%%%%%%%%%%%%%%%%%%%%%%
%%                                          %%
%% The keywords begin here                  %%
%%                                          %%
%% Put each keyword in separate \kwd{}.     %%
%%                                          %%
%%%%%%%%%%%%%%%%%%%%%%%%%%%%%%%%%%%%%%%%%%%%%%


% MSC classifications codes, if any
%\begin{keyword}[class=AMS]
%\kwd[Primary ]{}
%\kwd{}
%\kwd[; secondary ]{}
%\end{keyword}

\end{abstractbox}
%
%\end{fmbox}% uncomment this for twcolumn layout

\end{frontmatter}

%%%%%%%%%%%%%%%%%%%%%%%%%%%%%%%%%%%%%%%%%%%%%%
%%                                          %%
%% The Main Body begins here                %%
%%                                          %%
%% Please refer to the instructions for     %%
%% authors on:                              %%
%% http://www.biomedcentral.com/info/authors%%
%% and include the section headings         %%
%% accordingly for your article type.       %%
%%                                          %%
%% See the Results and Discussion section   %%
%% for details on how to create sub-sections%%
%%                                          %%
%% use \cite{...} to cite references        %%
%%  \cite{koon} and                         %%
%%  \cite{oreg,khar,zvai,xjon,schn,pond}    %%
%%  \nocite{smith,marg,hunn,advi,koha,mouse}%%
%%                                          %%
%%%%%%%%%%%%%%%%%%%%%%%%%%%%%%%%%%%%%%%%%%%%%%

%%%%% Note on publication
Accepted for publication in EPJ Data Science. 

%%%%% Introduction
\section{Introduction}

Almost 20 years after its birth, Wikipedia has become the reference  for  online diffusion of free encyclopedic knowledge, reaching $54$M articles in $313$ language editions\footnote{List of Wikipedias. \url{https://meta.wikimedia.org/wiki/List_of_Wikipedias}. Accessed March 2021.}. Its content is generated by the collaborative effort of a large community of editors, and provides a reliable source of information for web users~\cite{anthony2009reputation}. Knowledge on Wikipedia is mainly conveyed under the form of written text, but also through other types of content, such as references and images.

\begin{figure}[t]
  \centering
  \includegraphics[width=.95\linewidth]{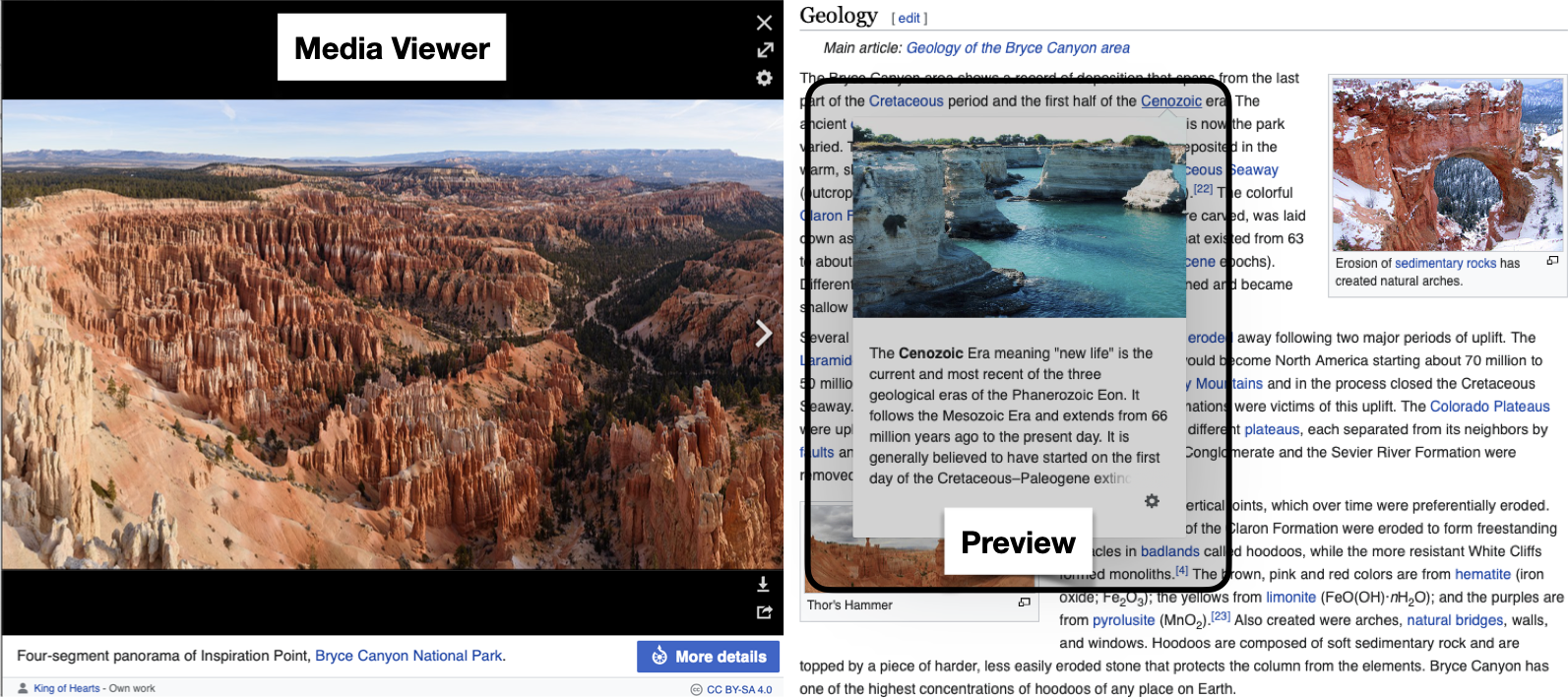}
  \caption{Examples of the two types of images visualization that we investigate in this study. The Media Viewer is opened when images are clicked, while Page Previews are shown when the reader hovers over a link to another page.}
  \label{fig:mediaviewer}
\end{figure}

The space of visual content on Wikipedia is vast. English Wikipedia alone contains more than $5$M distinct images, the majority of which is hosted by Wikipedia’s sister project Wikimedia Commons\footnote{Wikimedia Commons. \url{https://commons.wikimedia.org/wiki/Main_Page}. Accessed March 2021.}, the world's largest free visual knowledge repository. The genesis of Wikipedia images involves not only the contribution of Wikipedia editors, but also the participation of visual content \textit{creators}. Visual content on Commons and Wikipedia originates from individual photographers, artists, web users and cultural institutions in the GLAM spaces\footnote{Galleries, Libraries, Archives, and Museums}, who actively release their works of art for free and public use.

Given the crucial role of Wikipedia as a central hub for knowledge sharing and learning, understanding how images are used on Wikipedia is particularly important. A vast body of literature in experimental psychology has shown the impact of images for learning and engaging with knowledge. Images positively affect comprehension and increase attention on the textual material~\cite{levie1982effects}.
Despite their importance, while other aspects of Wikipedia have been widely studied~\cite{yasseri2012dynamics,lemmerich_why_2019,halfaker2020ores}, little is known about visual content and its usage, with only a few studies looking at cross-language image diversity~\cite{he2018the_tower_of_babel}, and the communities of Wikipedia ``image'' editors~\cite{viegas2007visual}.  

In this paper, we fill this gap in the literature by providing for the first time a comprehensive overview of how readers interact with images in (English) Wikipedia. We quantify and characterize reader engagement with images when browsing the encyclopedia using traffic data and we explore the role played by images in the exploration of free knowledge. 
To operationalize reader engagement, we adopt the most widely-used metrics in web user studies~\cite{bakhshi2014faces,park2015large}: we compute click-through rate on images, and conversion rate on illustrated and unillustrated page previews. While only partially representing the complex, multifaceted notion of interest~\cite{constantin2019computational}, these implicit signals do reflect an expression of engagement with visual content and they provide a solid baseline for an initial overview of readers' interactions with images. More specifically, we address three major research questions:

\begin{description}
    \item[RQ1:] \textbf{To what extent are readers interacting with images on Wikipedia?} And what is the relation with engagement values on other types of content? 
    \item[RQ2:] \textbf{What drives reader's engagement with images when reading Wikipedia articles?} What are the visual and contextual factors that influence image interactions? 
    \item[RQ3:] \textbf{Do images support reader's need for additional information when navigating Wikipedia?} Are images helpful to delve into contextual information provided by the article?
\end{description}

In addressing these questions, we make the following contributions:

\begin{itemize}
    \item \textbf{RQ1:} We collect a large dataset of reader interactions with images in English Wikipedia over one month and characterize the landscape of Wikipedia images with several features inspired by experimental psychology and web user studies (Sec. \ref{sec:image_features}). We quantify reader engagement with images and find that, on average, readers click with images $1$ in every $29$ pageviews on English Wikipedia, ten times more often than with references (RQ1, Sec. \ref{sec:rq1}).
    \item \textbf{RQ2:} To visualize the factors impacting reader engagement with Wikipedia images, we perform a set of multivariate analyses on the image features extracted and find that readers interact more often with images of monuments, maps, vehicles, and unfamiliar faces (RQ2, Sec. \ref{sec:rq2}).
    \item \textbf{RQ3:} To understand whether images support readers' need for additional contextual information when navigating Wikipedia, we design a matched observational study based on page previews, i.e., the short article summaries that are displayed when users hover on links to other Wikipedia pages (RQ3, Sec. \ref{sec:rq3}). We find a negative effect of the presence of images on the proportion of articles' page     previews that convert into a visualization of the full article page. 
\end{itemize}

We conclude  (Sec. \ref{sec:conclusion}) that the visual preferences of Wikipedia readers are radically different compared to web users in photo-sharing platforms or image search engines, where images of people and celebrities largely predominate. We also find that images on Wikipedia appear to fulfill part of the \textit{cognitive} function typical of illustrations in instructional settings supporting readers' information need. Finally, we discuss theoretical implication of this research and its important repercussions on how the Wikipedia communities organize and prioritize the inclusion of visual content and how the broader web and content creators could contribute to the web with free visual knowledge.

%%%%% Related Work
\section{Related work}
Our work is highly related to research from experimental psychology, computer vision, information retrieval, and computational social science, looking at how readers navigate knowledge.

\subsection{The role of text illustrations}
A substantial body of literature from experimental and educational psychology studied the role of images for knowledge understanding and learning.
Researchers have found that, very often, images in association with text help to support learning in instructional contexts \cite{guo2020you}, also in online settings \cite{mayer2002multimedia,khamparia2018impact,rudolph2017cognitive,tempelman2006multimedia}, especially when images are carefully curated, described, and positioned in the text~\cite{peeck1993increasing,bernard1990using}. Beyond this \textit{cognitive} purpose of facilitating content comprehension and providing complementary information, textual illustrations can have many other functions: the \textit{attentional} function, meaning that images can help to attract attention to the information in the textual form; the \textit{affective} function -- images help enhance emotions and enjoyment when reading a text; and the \textit{compensatory} purpose of supporting poor readers~\cite{levie1982effects}. 
While testing the role of images for knowledge understanding is beyond the scope of this paper, we borrow some ideas from these works to analyze reader interactions with images and design features and experiments aimed at replicating some of their findings.

\subsection{Image interestingness}
Several studies in computer science have looked at what makes images interesting from a computational perspective. Researchers have typically described interestingness in two ways. \textit{Visual} interestingness is the extent to which an image can hold or catch the viewer's attention due to its intrinsic visual qualities (see Constantin et al.~\cite{constantin2019computational} for a review of the most recent works in this space). Researchers have found that, for example, images are more interesting when they are more aesthetically pleasing or when their content is visually complex or unfamiliar~\cite{gygli2013interestingness}. \textit{Social} interestingness is often also called \textit{popularity} and corresponds to the extent to which an image is liked by a large number of people in a community. Social interestingness depends on the social dynamics of the platforms where images are shared, the pictures visual content~\cite{khosla2014makes,ding2019intrinsic,bakhshi2014faces} and the text associated with them~\cite{zhang2018user}.
Most of these previous works focus on predicting image popularity in photo sharing platforms such as Flickr~\cite{zhang2018user,khosla2014makes} or Instagram~\cite{bakhshi2014faces}, specifically designed to increase social interestingness in images. Unlike these existing works, we analyze here for the first time how readers engage with images in the context of online free knowledge spaces. We model the complex interplay between encyclopedic knowledge, pictorial representations, and reader engagement and explore the role of images informational support for Wikipedia articles.

\subsection{Image search behavior}
Related works have investigated web user behavior in image search engines. Researchers have found that, in general, the most popular queries in image search engines are about people, celebrities, and entertainment~\cite{tsikrika2014multi,jansen2008searching,huang2009analyzing}. By comparing image search behavior with text search behavior, several studies have found that image search sessions are heavier in interaction and exploration than the more ``focused'' textual sessions~\cite{jansen2004effect}, although, in a later study, O'Hare et al. found that web image search behavior is nonuniform across query types~\cite{park2015large}. While the scope of this work is different from this body of research, we will factor into our analysis findings from this area.

\subsection{Images on Wikipedia}
Recent works have quantified the monetary value and underlined the social contribution of Wikipedia's visual side \cite{heald2015valuation, erickson2018commons}. However, despite their important role, the space of images on the encyclopedia has rarely been investigated. Given the richness of their semantics, researchers have worked on building structured datasets from images in Wikimedia Commons, and Wikipedia 
\cite{vaidya2015dbpedia,ferrada2017imgpedia}. However, only a few works have focused on understanding editing behavior concerning images.  
A seminal study looked at understanding communities of editors who curate the visual content of the encyclopedia~\cite{viegas2007visual}. More recently, He et al.~\cite{he2018the_tower_of_babel} measured the visual diversity of Wikipedia, finding that cross-language image diversity is higher than the diversity of textual content and that many images are unique to specific language editions. Moreover, Navarrete et al. \cite{navarrete2020image} investigated the role of image paintings on Wikipedia, finding that they are extensively used to illustrate also non-art-related topic and that their audience is even larger than that of art-related articles. While most of these works focus on the image \textit{content} or the Wikipedia \textit{editor} communities, we study here the complementary aspect of how \textit{readers} interact with the visual content.

\subsection{Studying Wikipedia readers}
A few studies have focused on Wikipedia reader behavior, including reader article topic preferences~\cite{10.1145/2631775.2631805,spoerri_what_2007}, reader perception of the site performances~\cite{salutari2019large}, or reader informational need~\cite{lemmerich_why_2019,singer_why_2017}. More recently, Piccardi et al. worked on quantifying reader engagement with citations~\cite{piccardi2020quantifying} and external links~\cite{piccardi2021value}. The authors collected a large dataset of reader interactions with footnotes and references and showed that only a tiny portion of readers engage with citations on Wikipedia. In this direction, this work gives an additional perspective on Wikipedia readers' behavior, focusing on the volume and characteristics of reader interactions with visual content.

\subsection{User engagement metrics}
To quantify user engagement with images when reading Wikipedia, we borrow metrics used by several studies in the computational advertising field and user engagement studies~\cite{10.1145/2631775.2631805}. While these works aim to predict engagement metrics such as conversion rate~\cite{chapelle2014modeling,rosales2012post}, namely the percentage of landing page visits that result in a target action, or the click-through rate~\cite{ta2015factorization,richardson2007predicting,edizel2017deep}, namely the ratio between clicks and impressions, we use these metrics here as a means to decode Wikipedia reader behavior.

%%%%% Images on Wikipedia
\section{Images on Wikipedia}\label{wikiimages}
Images are a core component to help readers interpret knowledge in the encyclopedia and complement the textual information on Wikipedia articles. As English Wikipedia guidelines put it,
``The purpose of an image is to increase reader understanding of the article's subject matter, usually by directly depicting people, things, activities, and concepts described in the article.''\footnote{Image use policy. \url{https://en.wikipedia.org/wiki/Wikipedia:Image_use_policy\#Image_content_and_selection}. Accessed March 2021.}

Images are added to Wikipedia by hundreds of thousands of editors from all around the world, following a Manual of Style maintained by the Wikipedia community\footnote{Manual of Style. \url{https://en.wikipedia.org/wiki/Wikipedia:Manual_of_Style/Images}. Accessed March 2021.}. In essence, images added to Wikipedia articles have to be relevant to the article's content and of high photographic quality. The majority of images in the encyclopedia are hosted in Wikimedia Commons, the largest free visual knowledge repository. Images in Wikimedia Commons must be either of the public domain or licensed under a free license allowing anyone to reuse the material for any purpose, including commercial purposes.

\begin{figure}[t]
  \includegraphics[width=.95\linewidth]{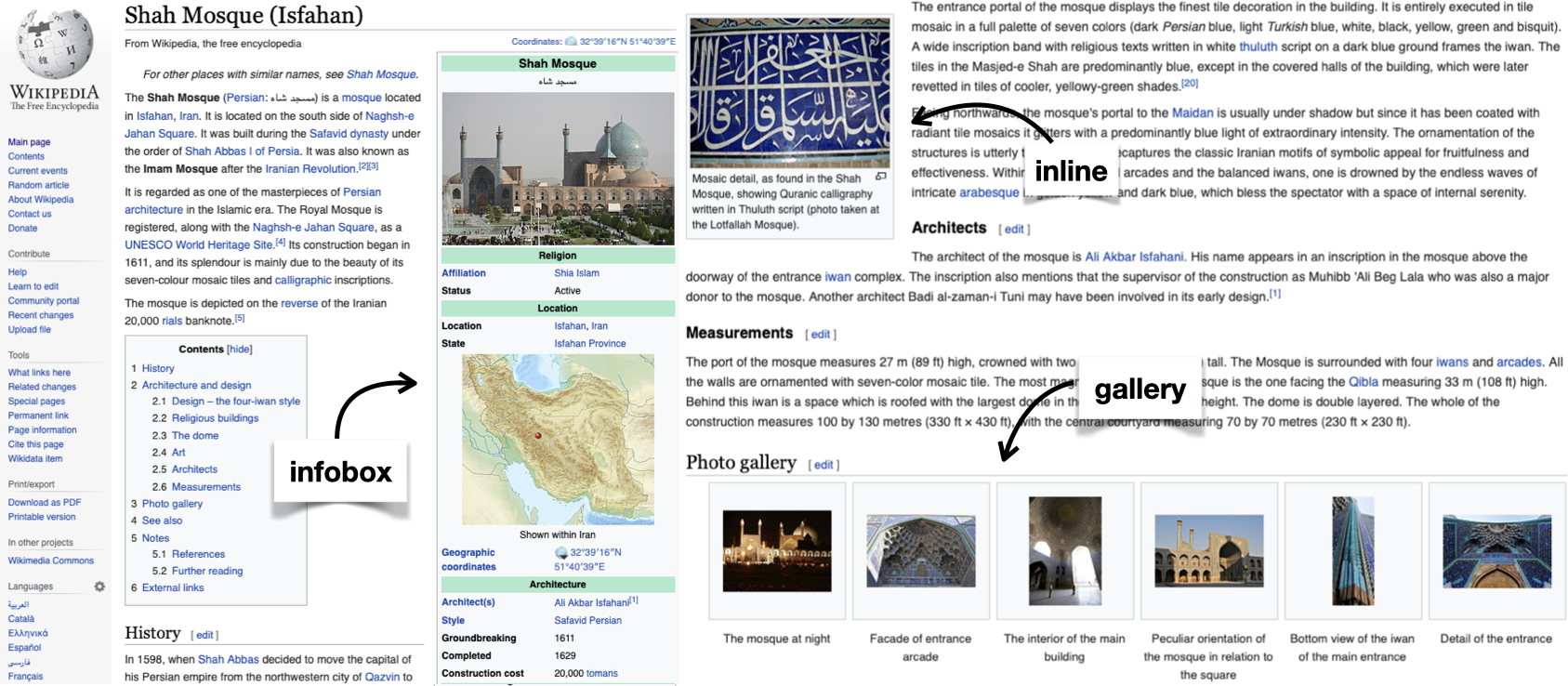}
  \caption{Examples of the three types of positions (\textit{infobox}, \textit{inline}, and \textit{gallery}) of images within a Wikipedia article.}
\label{fig:image_placement}
\end{figure}

Images can be placed in different parts of an article (see Fig. \ref{fig:mediaviewer} and \ref{fig:image_placement}). Readers can find images in the \textit{infobox}, a table summarizing the main facts about the article's subject, usually placed in the top-right corner of the page on desktop browsers or at the top of the screen in mobile browsers. Images can also be added \textit{inline}, namely individually near the relevant text in the article body. Finally, when images are too many to be placed within the text body, they can be collected into \textit{galleries}, generally added at the bottom of the articles. On desktop devices, images are also available in article \textit{previews}, i.e., the pop-up containing the article summary and an image (when available) which gets displayed whenever a reader hovers over on a link to another Wikipedia article.

For readers and editors who are interested in exploring Wikipedia visual content in further detail, images in articles are \textit{clickable}: when clicked, an image is previewed in a visualization tool called Media Viewer\footnote{The Media Viewer. \url{https://en.wikipedia.org/wiki/Wikipedia:Media_Viewer}. Accessed March 2021.}. The Media Viewer overlays on the article and displays the image in a larger size, and additional metadata below.

\begin{figure}[t]
  \centering
  \includegraphics[width=.95\linewidth]{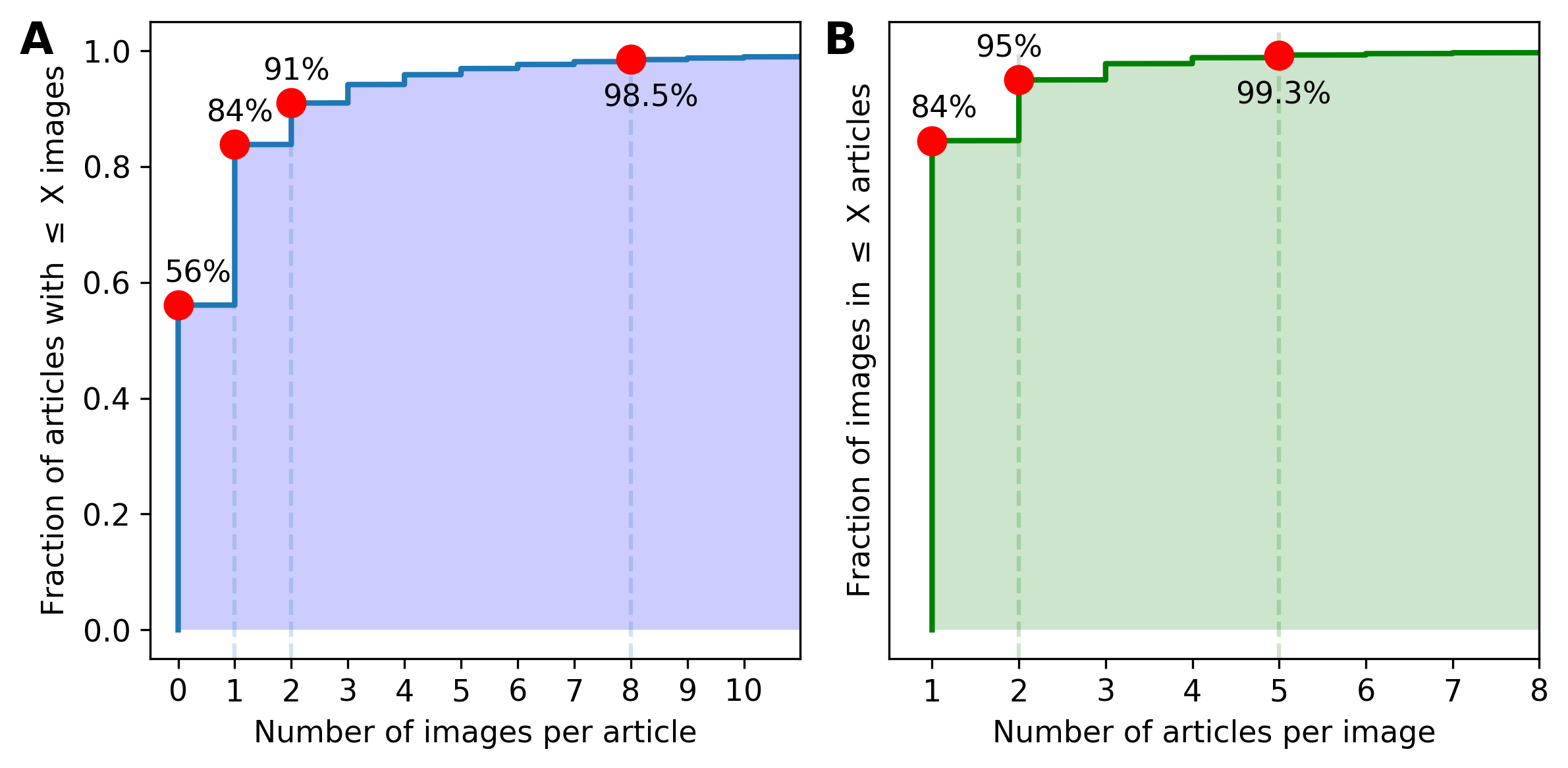}
  \caption{Cumulative distributions of (A) number of images per article and (B) the number of articles per image.}
  \label{fig:images_distribution}
\end{figure}

But how much visual content is available for readers to explore? If we take English Wikipedia, the largest language edition of the online encyclopedia, as of March 2021, it counted $6.2$M articles, for a total of $5$M unique images. As many other quantities on the Web, the distribution of images across Wikipedia articles follows a power law: as shown in Fig. \ref{fig:images_distribution}A, only $\approx44\%$ of pages in English Wikipedia are illustrated.

One reason for the number of missing images is the effort needed to illustrate articles. Wikipedia editors need to find the right image match for an article by searching through millions of images in Wikimedia Commons. However, when relevant images are not present in Commons, editors will have to search other sources. First, the right pictures for an article need to \textit{exist} somewhere on the Web (or in the world): otherwise, someone, Wikipedia editors, photographers, GLAM institutions, or other users, must create or retrieve them. Second, images have to be free to reuse. If images are not free-licensed, editors' and authors' efforts will be needed to make them publicly available. Only then images can be hosted in Wikimedia Commons and finally added to Wikipedia articles. 
To help with these efforts, the Wikimedia movement organizes several initiatives, e.g., Wiki Loves Monuments, encouraging photographers to add free images of monuments\footnote{Wiki Loves Monument. \url{https://www.wikilovesmonuments.org/}. Accessed March 2021.}, or the \#WPWP campaign, which helps editors add images to unillustrated Wikipedia articles\footnote{Wikipedia Pages Wanting Photos. \url{https://meta.wikimedia.org/wiki/Wikipedia_Pages_Wanting_Photos}. Accessed March 2021.}.

Given the central role of Wikipedia and its diverse content nature, knowing whether and how readers use visual information could help prioritize efforts around the visual enrichment of Wikipedia. 

%%%%% Data & Methods
\section{Data collection and methods}
To answer our research questions, we first need to estimate the volume of Wikipedia articles and their images, collect data about reader interactions with those, and characterize them through feature extraction.

\subsection{Collecting article and image counts data}
To measure the number of articles and images, we used the HTML version of English Wikipedia at the end of March 2021. We collected $6.2$M documents, and we parsed them to extract the images' URLs, caption text, resolution, and position on the page.
Using the CSS class in the HTML code, we exclude all images that appear as icons (for example, portals or Wikiprojects). Additionally, for each page, we also record the article length as the number of characters. 

Out of the $6.2$M articles, $2.7$M ($44\%$) contained at least one image, for a total of $5$M unique images across all English Wikipedia articles.
The vast majority of the articles ($91\%$) contain two images or less, while only $1.5\%$ has more than eight images (see Fig. \ref{fig:images_distribution}A). On average, there are $2.3$ images per illustrated article. Around $84\%$ of images is unique to the article where it appears, while $16\%$ of the images appear in more than one article (see Fig. \ref{fig:images_distribution}B).

\subsection{Collecting article and image traffic data}
\label{sec:traffic_data}
We obtained the reader interactions with images for desktop and mobile browsers by processing the server access logs\footnote{The Webrequest table. \url{https://wikitech.wikimedia.org/wiki/Analytics/Data_Lake/Traffic/Webrequest}. Accessed March 2021.} collected from 1st to 28th of March 2021. We restricted our analysis to only human interactions by ignoring traffic from bots thanks to a set of heuristics developed by Wikimedia's Analytics team\footnote{Bot or Not? Identifying “fake” traffic on Wikipedia, Wikimedia Analytics team. \url{https://techblog.wikimedia.org/2020/10/05/bot-or-not-identifying-fake-traffic-on-wikipedia/}. Accessed March 2021.}. For privacy reasons, we worked with an anonymized version without sensitive information. Since the logs do not contain any explicit identifier for the user, before the anonymization, we assigned a random id based on IP and user-agent similar to previous work~\cite{lemmerich2019world}. In addition, we discarded all the events coming from logged-in users, the events of any user that edited a page, and the events originated from countries where not all days have more than $500$ pageviews consistently. This filtering ensures more privacy for the Wikipedia readers by dropping around $3\%$ of the data.

Over the considered period, we selected from the web logs all requests that reflect three types of actions:
\begin{itemize}
    \item \textbf{Imageviews:} these requests correspond to image visualizations in the Media Viewer after a user clicks on an article image.
    \item \textbf{Pageviews:} these are requests logged every time a user visits a Wikipedia page. For the scope of this study, we select only pageviews of articles with at least one image.
    \item \textbf{Page previews:} these requests are logged whenever a user hovers over a link to an article. To remove the effect of casually generated page previews, we only keep those previews that are shown for at least one second. Note that page previews are generated only on desktop devices.
\end{itemize}

We aggregated these image-related events at the user level by using the previously assigned id to obtain sorted sequences of actions from the same user, which we refer to as \textit{sessions}.

In our analysis, we do not consider exogenous time-dependent events' impact and the role that external image search engines may play in directing users to Wikipedia. For these reasons, we filter out all the incoming traffic generated from Google Image Search, which represents by far the most used image retrieval engine from which people access Wikipedia's visual content. Nevertheless, the pageviews originating from Google Image search account for $0.006\%$ of the total, making their impact negligible.

\begin{figure}[t]
  \includegraphics[width=.95\linewidth]{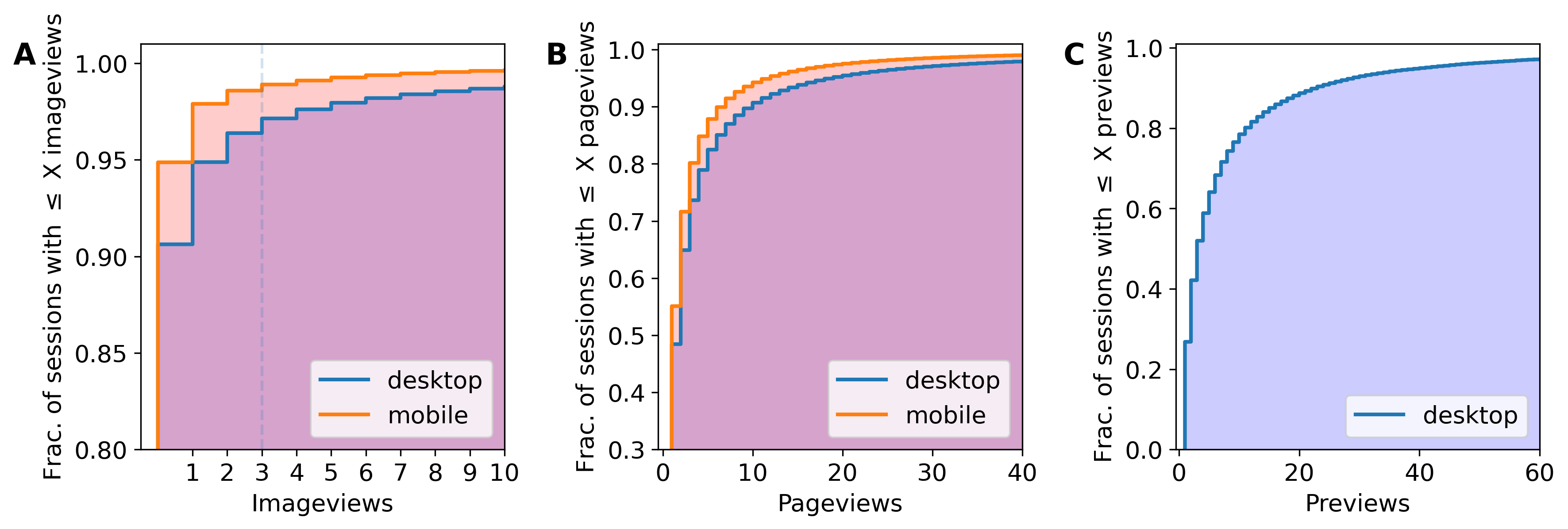}
  \caption{Cumulative distributions of number of sessions by (A) imageviews, (B) pageviews, and (C) previews partitioned by desktop (in blue) and mobile device (in orange). Page previews are available only on desktop devices.}
  \label{fig:views_distribution}
\end{figure}

In our data collection, we extracted interactions for $1.5$B sessions. In Fig. \ref{fig:views_distribution} we report the distributions of sessions by the number of imageviews, pageviews, and previews. The distributions are heavily skewed, with $91\%$ and $94\%$ of sessions having less than $10$ pageviews on desktop and mobile devices, respectively, and $99\%$ of sessions having less than $10$ imageviews both on desktop and mobile devices during our data collection period. Similarly, $79\%$ of sessions have generated less than 10 previews. Users with extensive sessions (i.e. ``power users''), that may be over-represented, are therefore limited in our analysis.
Over one month, $100\%$ of the illustrated articles have been loaded at least once, accounting for a total of $7.1$B pageviews, $461$M imageviews, and 49M previews events in our dataset. We find that most pageviews are generated from mobile devices ($59\%$ from the mobile site), while most imageviews are generated from desktop ($58\%$ from desktop).

\subsection{Mining image content and context}
\label{sec:image_features}
To investigate the factors that make images engaging when reading Wikipedia, we characterize the pictures in our dataset with several features related to the visual context and content. Our choice of features is largely inspired by the literature around the cognitive perception of images in instructional or web environments.

\subsubsection{Contextual factors}
Images on Wikipedia are not isolated items. Instead, they exist in \textit{context}, providing epistemic support to the article they are illustrating. To extract features from the image context, we resort to previous literature on Wikipedia reader behavior and experimental psychology studies on the role of images in instructional settings. 
Note that $16\%$ of the images appear in multiple articles. Since the same image may appear in very different articles, thus belonging to very different contexts, we treat such images as distinct.

\begin{figure}[t]
 \includegraphics[width=.95\linewidth]{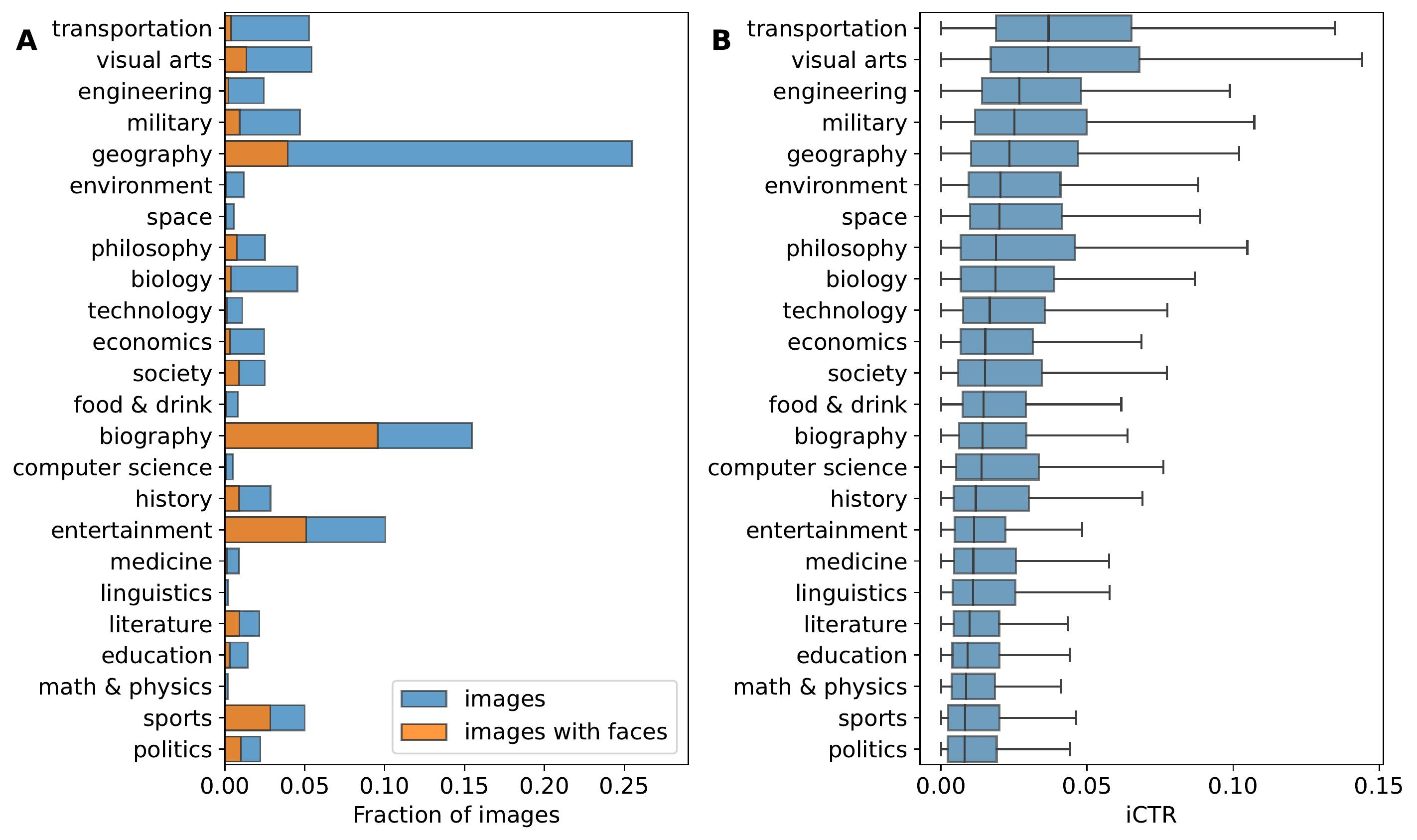}
  \caption{(A) Fraction of images by topic (in blue) and fraction of images with faces (in orange). (B) Image-specific CTR by article topic.}
  \label{fig:topic_distributions}
\end{figure}

\paragraph{Page topic} In a previous study, Piccardi et al.~\cite{piccardi2020quantifying} found that Wikipedia reader engagement with references varies with the article topic. To test whether the reader's need for visual support similarly varies across subject matters, we extract, for each page in our dataset, a \textit{topic vector}, using the Wikidata topic model\footnote{Wikidata topic model. \url{https://github.com/geohci/wikidata-topic-model}. Accessed March 2021.}. The classifier takes as input the Wikidata item of a Wikipedia page, and it returns a $64$-dimensional vector containing the probability that the article belongs to the topics of the Wikiproject hierarchy\footnote{The WikiProject Directory. \url{https://en.wikipedia.org/wiki/Wikipedia:WikiProject\_Directory}. Accessed March 2021.}. To reduce the dimensionality of the topic vector,  we consider the second level of the topic taxonomy accounting for $31$ topics. We then rearranged some of the topics into coarse-grained topics, namely media, internet culture, and performing arts into \textit{entertainment}, chemistry and biology into \textit{biology}, computing and libraries \& information into \textit{computer science}, mathematics and physics into \textit{maths \& physics}. 
Fig. \ref{fig:topic_distributions}A shows the distribution of images by article topic. Geographic articles are the most illustrated, containing $1/4$ of the images in our dataset. Biographies, making up $30\%$ of the articles on Wikipedia, also contain around $15\%$ of the images. Topics such as entertainment (movies, plays, books), visual arts, transportation, military, biology, and sports follow, covering together another third of the images in English Wikipedia. A summary of the numerical values can be found in the Supplementary Material (Supplementary Table).

\begin{figure*}[t]
  \centering
  \includegraphics[width=0.95\linewidth]{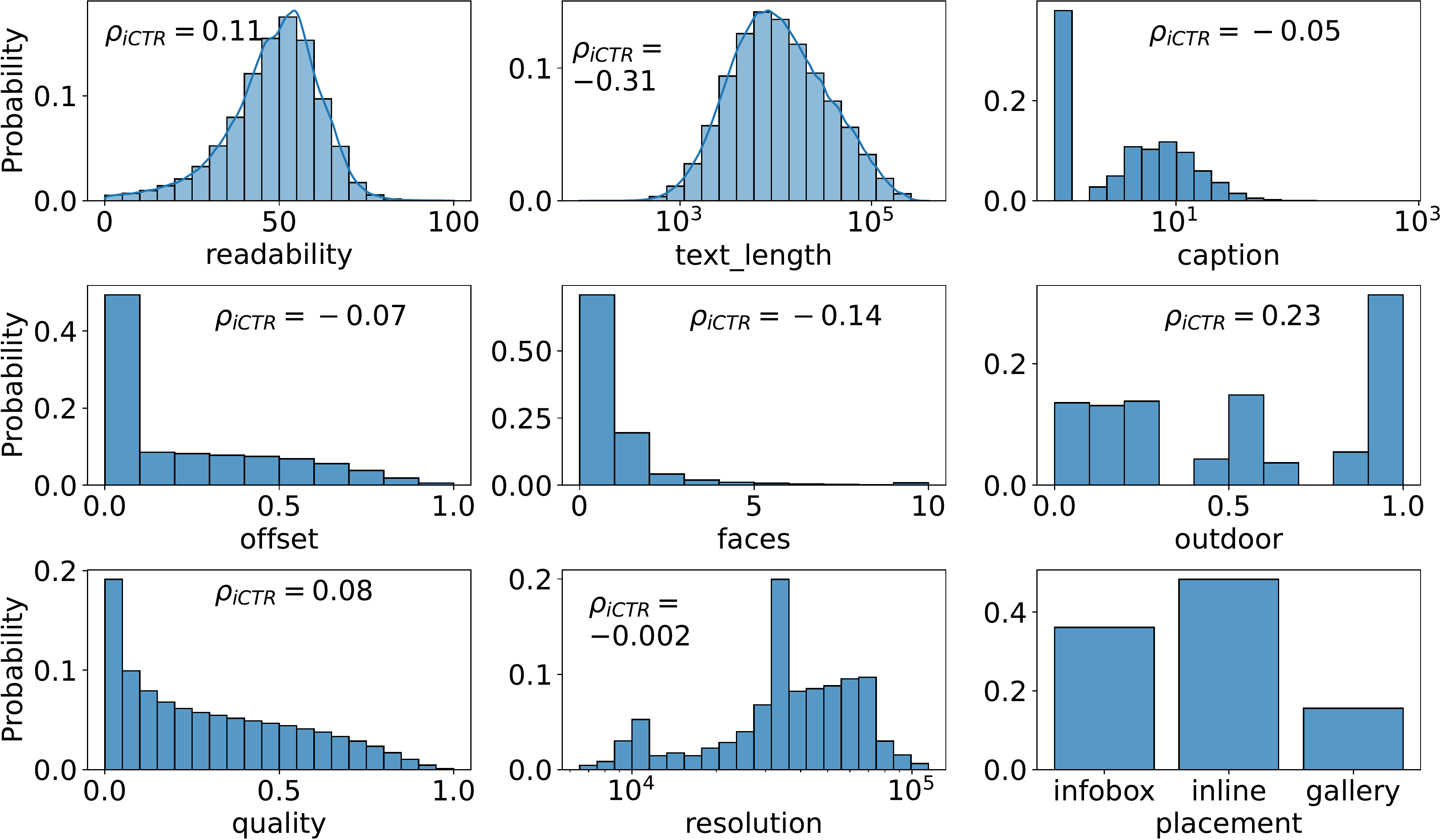}
  \caption{Feature distributions. Spearman's rank correlation coefficient $\rho$ between the numerical features and the iCTR at the top of each panel ($p<0.001$ for each feature).}
  \label{fig:feature_distributions}
\end{figure*}

\paragraph{Page length} One of the possible functions of text illustrations in learning contexts is to enrich and complement the textual content with additional material~\cite{levie1982effects}. To investigate to which extent images are used to complement the lack of textual information, we measure the textual richness as the \textit{length} of each article in characters. The distribution of the number of images by text length as shown in Fig. \ref{fig:feature_distributions} is log-normal, with most images in English Wikipedia being found in articles between $1$k and $100$k characters long.

\paragraph{Page popularity} Previous work analyzing reader behavior with respect to Wikipedia citations~\cite{piccardi2020quantifying} found that there is an inverse relation between article popularity and reference click-through rate. To test whether this relation is valid also in the case of interactions with images, we compute a page \textit{popularity} feature for each image by computing the total monthly pageviews for the page where an image appears. As in Piccardi et al.~\cite{piccardi2020quantifying}, the page popularity follows a power-law distribution (Fig. \ref{fig:ictr_pageviews}), with $70\%$ images having average monthly pageviews in the range between $50$ and $10$k.

\paragraph{Readability} Although not completely verified, another function of images in textual knowledge is to facilitate text comprehension, especially in the case of reading difficulties~\cite{levie1982effects}. To take into account this function in our study, we quantify the reading ease by computing the Flesch \textit{readbility} score, reflecting the ``comprehension difficulty of written material''~\cite{flesch1948new}, on the text of each article in our dataset. We compute the readability score for all the pages containing an image, and plot the resulting distribution in Fig. \ref{fig:feature_distributions}: most of the images on Wikipedia are in articles detected as ``Fairly difficult to read'' (score $50$-$60$), or ``Difficult to read'' (score $30$-$50$).

\paragraph{Length of the image caption} Studies in educational technologies have found that the usage of captions marginally enhances the usefulness of text illustrations~\cite{bernard1990using}. To operationalize the presence of captions as a contextual feature of the images in our dataset, we store the average number of words used to caption each image when appearing in a Wikipedia article. We can see from Fig. \ref{fig:feature_distributions} the caption length following a Tweedie distribution with a large fraction of the images without a description and the majority of existing captions centered around ten words.

\paragraph{Image placement} How images are placed in the text can play a crucial role in the knowledge exploration experience~\cite{peeck1993increasing}, and researchers investigating Wikipedia reader behavior showed that people tend to engage more with content (in this case, internal hyperlinks) which lies at the top of the article~\cite{paranjape2016improving}. At the same time, Wikipedia editors follow specific placement guidelines when illustrating an article. To investigate the role of image placement on Wikipedia article consumption, we extract the image's \textit{text offset}, i.e., the relative position of the image with respect to the length of the article, as well as the image \textit{position}, a categorical feature which can take the values $\{infobox, inline, gallery\}$, depending on the template used to add the image to the article. From the plots in  Fig. \ref{fig:feature_distributions} we can see that only 36\% of the images in our dataset is generally placed in infoboxes, while only 16\% can be found in galleries, and that the majority of inline images are generally placed at the top of the article (see \textit{offset}). A summary of the numerical values can be found in the Supplementary Material (Supplementary Table).

\paragraph{Image resolution} In addition to their position, the viewer's attention may also be driven by the size of an image. According to the Wikipedia's Image Size guidelines\footnote{Wikipedia:Manual of Style/Images Size. \url{https://en.wikipedia.org/wiki/Wikipedia:Manual_of_Style/Images\#Size}. Accessed March 2021.}, editors should choose the appropriate image size in proportion of its level of details. However, readers may still tend to click on small images that are inherently difficult to observe. To investigate the role of the image size, we compute the image \textit{resolution} in pixels for each image. As shown in Fig. \ref{fig:feature_distributions}, image resolutions vary across different scales, mostly ranging from $10$k to $100$k pixels.

\subsubsection{Visual factors}
The content of pictures plays a key role in driving readers' attention to both the images ~\cite{gygli2013interestingness} and the text on the page~\cite{levie1982effects}. To understand the type of visuals that elicit higher levels of interactions with Wikipedia images, we run a set of computer vision-based classifiers. Since training a classifier to detect every concept in Wikipedia's visual knowledge would be practically infeasible, we instead focus on three main indicators, based on extensive literature from visual and social interestingness prediction.

\paragraph{Image quality} Visual aesthetics, or image quality, is one of the top visual factors driving the viewer's attention to an image~\cite{gygli2013interestingness}. At the same time, researchers have found that not all images which receive much attention from web communities are actually of high quality~\cite{schifanella2015image}, and that a lot of socially uninteresting pictures are very beautiful. We investigate here whether the quality of an image plays an important role in eliciting Wikipedia reader attention. To do so, we design a \textit{Wikipedia Image Quality} classifier, as follows.

\begin{itemize}
    \item We collect a training set of images annotated with a binary (high/low) image quality score. To annotate images, we resort to the highly curated categories that Wikimedia Commons editors assign to images. We download $141,984$ images from the \textit{Quality images} category from  Commons\footnote{Commons:Quality images. \url{https://commons.wikimedia.org/wiki/Commons:Quality_images}. Accessed March 2021.}: these are high-quality images that have to meet Commons' quality guidelines\footnote{Commons:Image guidelines. \url{https://commons.wikimedia.org/wiki/Commons:Image_guidelines. Accessed March 2021.}} before being voted and promoted as Quality images by the community through a highly selective process. Only a few images make it to the ``image quality" category: there is, therefore, a large consensus on the quality of the images in that category. To collect low-quality images, we simply randomly sample an approximately equal number of pictures ($169,310$) from the large pool of Commons images. These are very likely to be low quality, as images randomly drawn from Commons tend to have a small resolution, and they are rarely used to illustrate Wikipedia articles~\cite{erickson2018commons}.
    \item We next train a deep neural network using transfer learning: we fine-tune a pre-trained model, originally designed to classify image objects, using the image quality data collected. We use the Inception-v3~\cite{szegedy2016rethinking} deep network pre-trained on the $1000$-classes ImageNet dataset~\cite{deng2009imagenet}, as it was proved to be a good starting dataset for transfer learning tasks~\cite{huh2016makes}. We use $90\%$ of the data for training and the rest for validation, and we train the last layer of the network over 10,000 iterations with the data collected. The fine-tuned classifier achieves $77\%$ accuracy on a balanced test set.
\end{itemize}

The resulting image quality classifier, given any image, outputs a \textit{quality score} in the range $[0,1]$ which corresponds to the probability that the image belongs to the ``High Quality'' class. As shown in Fig. \ref{fig:feature_distributions}, most images in our dataset have a very low-quality score.

\paragraph{Presence of faces} In line with several studies showing the importance of faces for web users' positive reactions and engagement with images~\cite{bakhshi2014faces,pappas2016multilingual}, we also extract information about the presence of faces of people in the image. We use MTCNN~\cite{wen2016discriminative} to detect faces and their bounding box in an image. For a given image, we then output a binary feature indicating whether it contains at least one face or not.
We find that around $1/3$ of the images on Wikipedia have at least one face (see Fig.~\ref{fig:feature_distributions}), and most of those are in articles about biographies, entertainment, and sports (see Fig.~\ref{fig:topic_distributions}A).

\paragraph{Outdoor setting} Literature around image interestingness and aesthetics~\cite{gygli2013interestingness} has shown that outdoor images tend to elicit the viewer's interest more than indoor images do. To extract the information about the image scene setting, we use a Wide Residual Network~\cite{zagoruyko2016wide} trained on MIT's Places~\cite{zhou2017places}, an image dataset with 10M images annotated with 365 scene types, and indoor/outdoor labels. This classifier, given an image, outputs an \textit{outdoor} score which reflects the probability of the image being an outdoor scenery. When the feature is $\leq 0.5$, the image is likely to be an indoor scenery. In our dataset, indoor and outdoor images are almost equally distributed, with a slight prevalence of outdoor pictures.

\subsection{Engagement metrics}
To quantify the volume of readers' interactions with visual content, we introduce the following metrics:

\paragraph{Global click-through rate.} The global click-through rate (gCTR) measures the overall reader engagement with images. It is defined as the fraction of reading sessions with at least one interaction with an image. Formally, for each session $s$, let $C(s,p)$ be the indicator function that is $1$ if at least one image was clicked on page $p$ by the respective reader, and $0$ otherwise. Moreover, let $N(p)$ be the number of distinct reading sessions during which page $p$ was loaded. We define the global click-through rate as

\begin{equation}
    gCTR = \frac{\sum_s \sum_p C(s,p)}{\sum_p N(p)}
\end{equation}

where p ranges over the set of pages that contain at least one image.

\paragraph{Image-specific click-through rate.} The image-specific click-through rate (iCTR) measures how much engagement a Wikipedia image elicits. It is defined as the ratio of clicks to impressions. Formally, let $N(i)$ be the number of distinct sessions with clicks on image $i$ and $N(p_i,i)$ the number of distinct sessions that viewed page $p_i$ where the image is placed, the image-specific click-through rate is
\begin{equation}
    iCTR(i) = \frac{N(i)}{\sum_{p_i \in P_i} N(p_i,i)}
\end{equation}
where $p_i$ ranges over the set $P_i$ of pages containing $i$.

\paragraph{Conversion rate.} The conversion rate (CR) quantifies the probability of clicking on an article link after its preview is shown in another article. Formally, for each page $p$ and session $s$, we denote by ${C(s,p)}$ the indicator function that is one if session $s$ has clicked on a link to page $p$ after seeing its preview. Moreover, we denote by $N(p)$ the total number of distinct sessions that loaded a preview of $p$. The conversion rate for page $p$ can be written as:

\begin{equation}
    CR(p) = \frac{\sum_s C(s,p)}{N(p)}
\end{equation}

In the following sections, we restrict our analyses to images visualized by at least $50$ readers during the period of our data collection in order to reduce the effect of rarely viewed articles and obtain a reliable estimate of the quantities above. This results in a set of $3.2$M unique images displayed in $2.7$M articles.

%%%%% RQ1
\section{RQ1: To what extent are readers interacting with images in Wikipedia?}
\label{sec:rq1}
The first step of our analysis is to quantify the volume of readers' interactions towards visual content when reading Wikipedia. To this aim, we compute the global click-through rate and image-specific click-through rate on our data and find the following.

\subsection{Overall engagement with images: the global click-through rate}
We find that the gCTR across all pages in English Wikipedia with at least one image is $3.5$\%, meaning that around $3.5$ out of $100$ times readers visit a page, they also click on an image. This metric is higher for desktop ($5.0\%$) and lower for mobile web users ($2.6\%$), probably due to differences in the way readers navigate Wikipedia on the two devices and the better Media Viewer experience on desktop.
Over time, the behavior also changes depending on the device used. For example, on desktop, readers tend to click more often on images during weekdays (Monday to Friday), with an increase of $5.5\%$ over weekends. However, on mobile, there is no significant difference between week and weekends.
To understand whether these values represent a high or low level of engagement, we can compare them with engagement metrics on another type of article content, namely article's references. According to Piccardi et al.~\cite{piccardi2020quantifying}, the gCTR on citations in English Wikipedia is $0.29\%$, thus around ten times lower than for images. This observation suggests that images tend to elicit a different level of engagement than those on references for English Wikipedia.

\begin{figure}[t]
  \includegraphics[width=.95\linewidth]{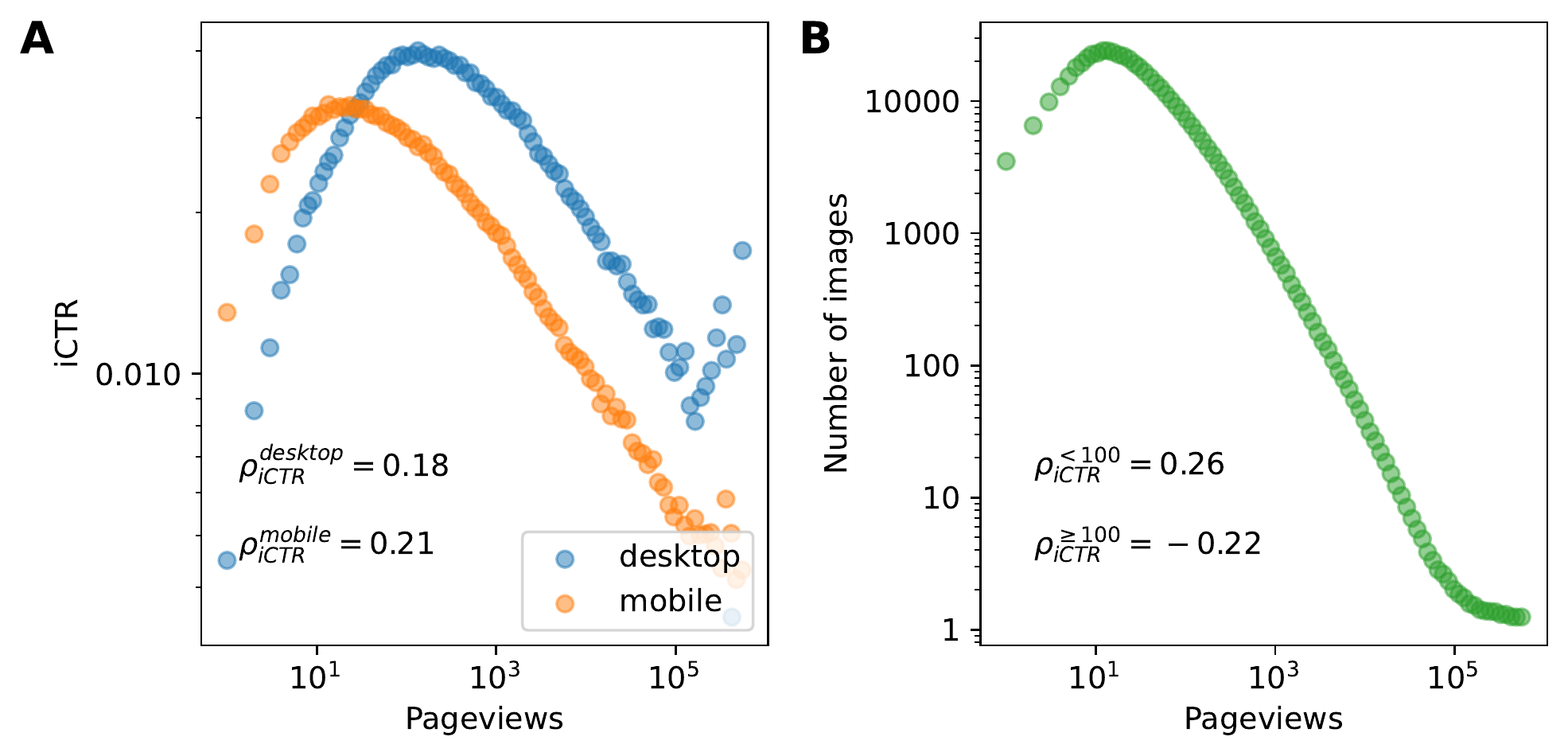}
  \caption{
  Distributions of (A) values of iCTR by page popularity partitioned by device and (B) number of images per page popularity. Spearman's rank correlation coefficient $\rho_{iCTR}$ between iCTR and pageviews in the inset ($p<0.001$). The axes are in log scale.}
  \label{fig:ictr_pageviews}
\end{figure}

\subsection{Average engagement with individual images: image-specific click-through rate} 
\begin{figure}[t]
  \includegraphics[width=.95\linewidth]{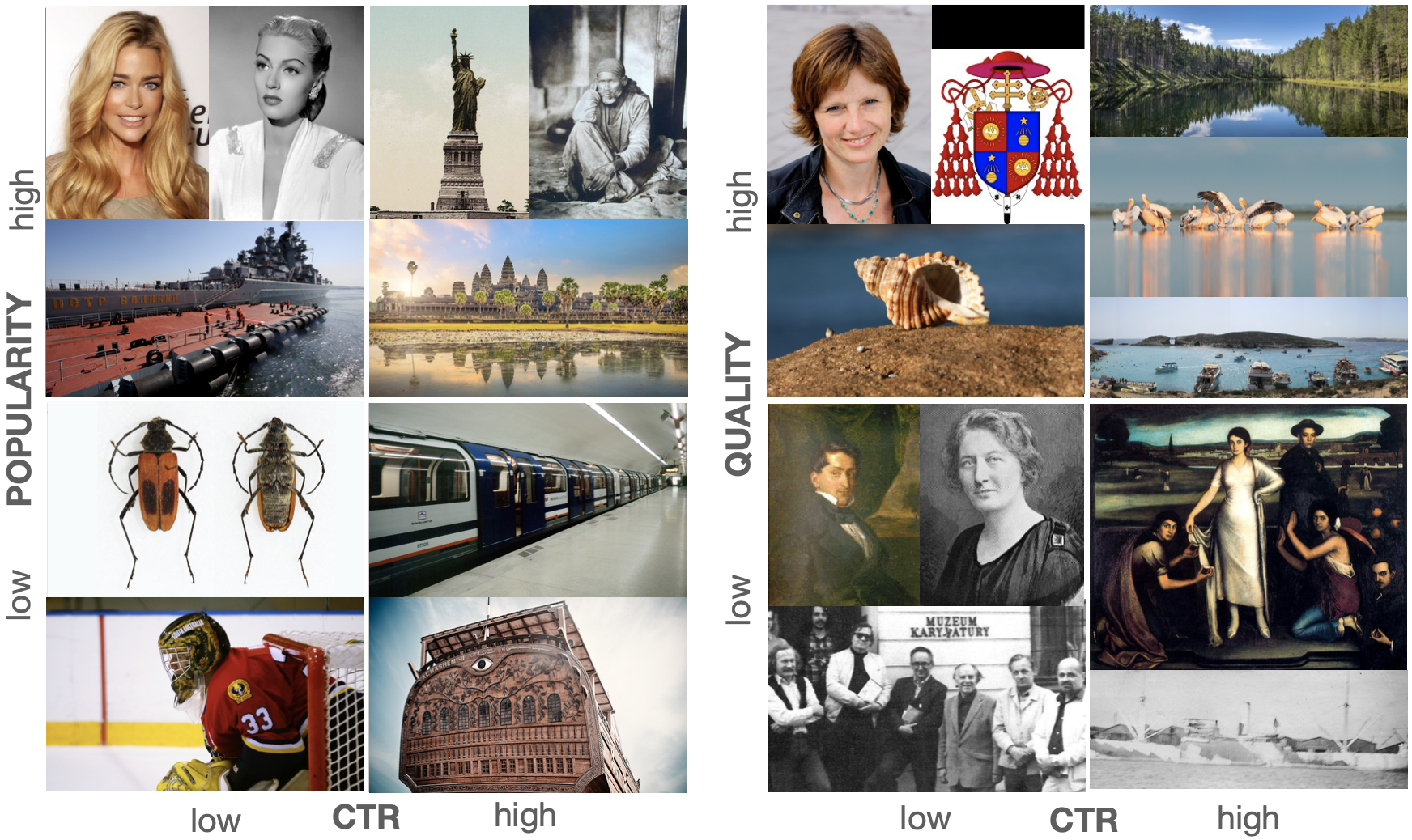}
  \caption{Examples of \textit{high} and \textit{low} image-specific CTR images by page popularity (left)  and image quality (right). We ranked images by iCTR, popularity and quality, and picked examples from the top-100 (``high'') and bottom-100  (``low'') for each dimension.}
  \label{fig:image_grid}
\end{figure}

On average, an image in a Wikipedia article gets clicked $2.6$ times every $100$ impressions. Again, the iCTR is higher ($3.2\%$) for desktop than for mobile users ($2.2\%$). In Fig. \ref{fig:image_grid} we report examples of highly engaging and less engaging images. By visually inspecting these results, we can see some visual trends: highly engaging images seem to depict outdoor environments. In contrast, among the images with low levels of iCTR, we can find human faces. 

%%%%% RQ2
\section{RQ2: What drives reader's engagement with images when reading Wikipedia articles?}
\label{sec:rq2} 
To address RQ2, we now model reader interaction with images on Wikipedia using the factors listed in Sec. \ref{sec:image_features}.

\subsection{Exploratory analysis}
\label{sec:exploratory_analysis}
We start our analysis by seeking a relationship between our target metric, the iCTR, and each of the contextual and visual factors in Sec. \ref{sec:image_features}. We report the Spearman's rank correlation coefficients $\rho_{ctr}$ between the iCTR and the scalar predictors in Fig. \ref{fig:feature_distributions} and \ref{fig:ictr_pageviews}. Considering the contextual factors, we observe a negative correlation with article length and popularity ($\rho=-0.31$ and $\rho=-0.21$, respectively). When further investigating the relationship with article popularity (Fig. \ref{fig:ictr_pageviews}), we find that it seems non-linear: engagement with images is low for highly unpopular pages. It becomes higher for pages in a mid-level bucket of popularity and drops again for highly viewed pages.
Regarding the image size, despide images are displayed in different resolutions, this does not have a clear relation with the iCTR ($\rho=-0.002$). When considering the position in the page instead, the median iCTR is higher for images in galleries (median iCTR=$0.024$) than for images in the infobox (median iCTR=$0.019$) and inline (median iCTR=$0.016$).
Moreover, we see signals of reader visual preferences in terms of article topics (Fig. \ref{fig:topic_distributions}B): the topics with the highest median value are transportation ($0.037$) and visual arts ($0.037$), while politics and sports show the lowest level of interaction with a median iCTR of $0.008$. 
Finally, the correlation analysis of the visual factors confirms our initial intuitions from the visual analysis. There is a positive correlation between the iCTR and outdoor scenery ($\rho=0.23$) and a negative relation between the presence of faces and readers' engagement ($\rho=-0.14$). A complete summary of the numerical values discussed can be found in the Supplementary Material (Supplementary Table).

\subsection{Regression analysis}
\label{sec:log_reg}
Next, we aim to understand how much these features are predictive of reader engagement with images. To do so, we perform a logistic regression analysis that classifies images according to their iCTR.

\begin{figure}[t]
  \includegraphics[width=0.95\linewidth]{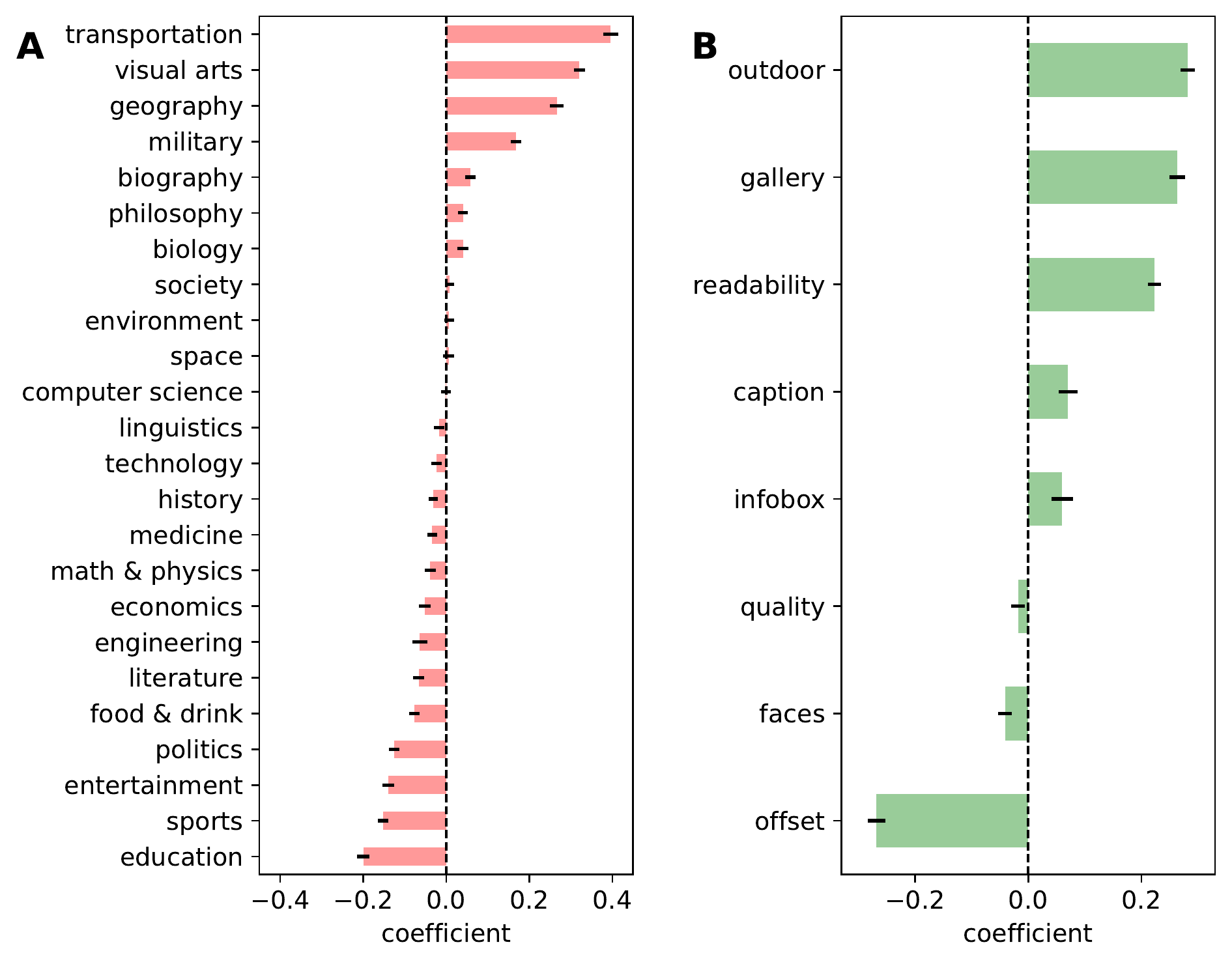}
  \caption{Association of the features with the image iCTR expressed as coefficients of the logistic regressions. (A) Coefficients of the model trained with topics of the article as predictors. (B) Coefficients of the model trained with the other variables of the image. Error bars represent 95\% confidence intervals.}
  \label{fig:logreg_coef}
\end{figure}

\paragraph{Study design}
We build the training set as follows. We take the median value of iCTR and label the images in our dataset with two classes of \textit{high} and \textit{low} iCTR according to whether their iCTR is above or below the median\footnote{We repeat the logistic regression analysis with different thresholds splitting the two classes, namely we focus on the highest vs. the lowest percentiles of the images according to their iCTR. We find no significant differences on the resulting regression coefficients. Therefore, we choose the median as the cutoff to maximize the presence of images in the analysis.}. We use the contextual and visual factors described in Sec. \ref{sec:image_features} as predictors and the binary iCTR as the target variable. Moreover, we split the predictors into two sets of features and train two separate logistic regression models. The first set of features consists of the topic vectors, while the second consists of the remaining other factors. In the second set of features, we log-transform variables that span over different scales, such as page popularity, text length, caption length, and the number of faces. Moreover, to reduce the amount of multicollinearity among the predictors, we manually inspect the correlation table and compute the Variance Inflation Factor~\cite{kutner2005applied} for each variable. We decide to exclude the \textit{inline} variable, as it shows strong collinearity with \textit{gallery} and \textit{infobox}. Finally, we standardize each predictor in the two sets of features.

\paragraph{Impact of image resolution}
We found that images on Wikipedia are displayed in different resolutions. Before running the regression analysis, we test the hypothesis that the image size could be decisive in attracting clicks,  i. e. readers may tend to click on smaller images as it may be harder to see the details. In Sec. \ref{sec:exploratory_analysis} we found the correlation coefficient to be $-0.002$ (with $p < 0.001$), indicating no clear relationship between the two variables. Moreover, we observe that image resolution is highly related to its position within the page: the median resolution is about 46, 36, and 11 megapixels respectively for images in the infobox, inline, and in galleries. Also, image resolution is highly correlated with some topics, e. g. it has large positive correlation with biography and entertainment, and large negative correlation with geography and visual arts. Since the image resolution does not seem to be directly related to the iCTR, while it seems to be influenced by some other independent variables, and thus may act as a confounder, we decide not to take it into account in the subsequent analyses.

\paragraph{Controlling for page length and popularity}
Similar to what was described in previous work on engagement with Wikipedia content~\cite{piccardi2020quantifying}, we found that the page popularity and the text length have strong negative correlations with the target. Since page popularity and text length show large variations across the other predictors, especially across topics, we remove the effect of these two confounding variables with a matched study. We build a bipartite graph with images of low and high iCTR as nodes of the two halves. We split the log-transformed page popularity and text length ranges into $100$ bins of equal size each, and assign the nodes to these bins, linking two nodes of opposite iCTR when falling into the same bins of popularity and length. Finally, we use min-weight matching on the bipartite graph to find pairs of high/low iCTR samples that minimize the Euclidean distance between all pairs. This procedure succesfully balanced the dataset, with the standardized mean difference of text length and pape popularity across the two classes dropping from $-0.54$ and $-0.51$ to $-0.010$ and $-0.007$, respectively.

\paragraph{Results}
The resulting regression models have an area under the ROC curve (AUC) of $0.67$ and $0.62$ for the model trained on the topics and the model trained on the other variables, respectively. Fig. \ref{fig:logreg_coef} shows the resulting models coefficients. In Fig. \ref{fig:logreg_coef}A, we observe that clicks on images are more often related to topics such as transportation, visual arts, geography, and military. On the contrary, clicks on images are less likely in education, sports, and entertainment articles. In Fig. \ref{fig:logreg_coef}B, we observe that the most important negative predictor is the text offset, i.e. the relative position of the image with respect to the length of the article, meaning that images are more clicked if placed in the upper part of an article. Regarding the visual content, we observe a strong positive effect of outdoor settings, consistently with the positive coefficients of transportation and geography, topics in which a large portion of images display outdoor scenes. Regarding the image position on the page, we find that images in galleries show a high level of engagement, as well as images in the infobox, even though with a moderate effect. Noteworthy, the presence of faces has negative impact in predicting a high level of interactions with images, contrary to what we would expect from the literature \cite{bakhshi2014faces}. In the remainder of this section, we further investigate this inconsistency in depth, by performing a clustering experiment and an observational study on the images in our dataset.

\subsection{Identifying prototypical image groups}
\label{sec:clustering}
To dive deeper into the results emerging from the regression analysis, we provide in this section a non-linear multivariate analysis of our data.

\paragraph{Study design}
Our goal is to draw a complementary picture of the complex interplay between reader engagement and image features, identifying prototypical groups of Wikipedia images with homogeneous characteristics. To this extent, we perform a density-based clustering using HDBSCAN~\cite{campello2013density}, which seeks partitions with high density areas of points separated by low density areas, possibly containing noise objects. 
The advantage of using HDBSCAN is threefold: first, its density-based structure allows to better identify areas of continuous, non-globular points compared to other clustering algorithms that rely on the assumptions of spherical shape clusters, e.g., k-means~\cite{lloyd1982kmeans}. Second, by labeling the sparse background points as noise, it aggregates data into coherent clusters rather than partitions. Finally, it extends DBSCAN~\cite{ester1996density} by implementing a hierarchical clustering approach that allows to extract the optimal flat grouping based on the stability of the clusters, allowing to find groups with non homogeneous density in contrast to a global density threshold adopted by DBSCAN.    

We run HDBSCAN\footnote{To run the algorithm, we use the \textit{hdbscan} Python library \cite{McInnes2017}: \url{https://hdbscan.readthedocs.io}.} on the features set described in Sec.~\ref{sec:log_reg} including the binary iCTR variable and limiting the analysis to the eight most popular topics (geography, biography, entertainment, visual arts, transportation, sports, military, and biology) that account for 92\% of the images in our corpus.
HDBSCAN has two main hyper-parameters that have significant practical effect on the clustering: \textit{min\_cluster\_size} which refers to the minimum number of grouped items to consider as a cluster, and \textit{min\_samples} which provides a measure of how conservative the clustering would be defining the level at which points are considered noise. The larger the value, the more conservative the clustering, that implies more points will be declared as noise, and clusters will be restricted to progressively more dense areas. 
We explore the hyper-parameter space with a grid search approach to find the best configuration that maximizes the Density-Based Clustering Validation (DBCV) index \cite{moulavi2014density}. 
Due to computational constraints, we perform the clustering on a random sample of 50K images, we repeat the procedure 5 times to assess the stability of the tuning phase. We achieve the best configuration with $min\_cluster\_size=600$ and $min\_samples=5$ in the majority of the runs.      
With these settings, we identify $23$ clusters, with a number of images ranging between 600 and 5000.

\begin{figure*}[t]
\includegraphics[width=.95
\linewidth]{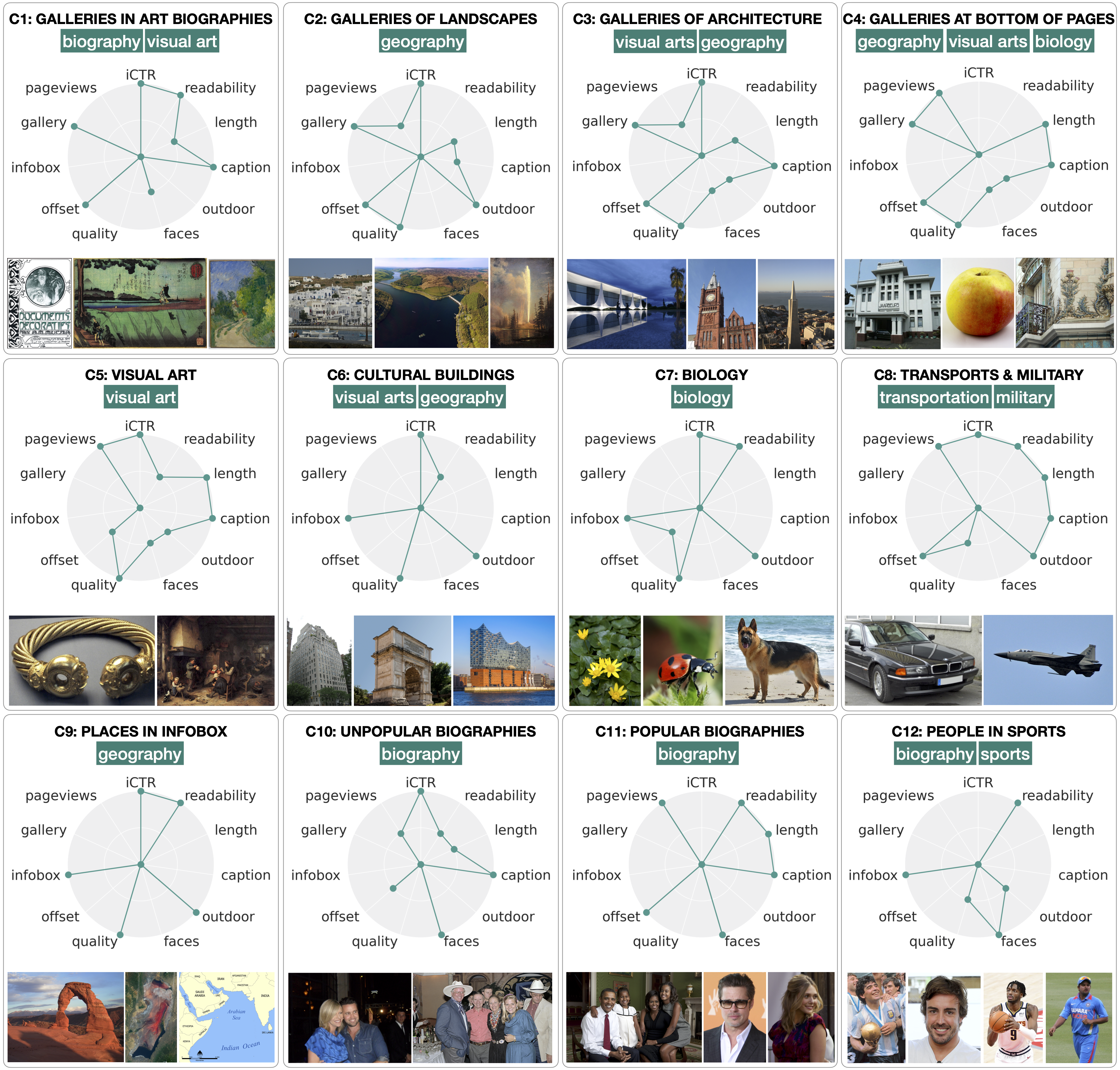}
\caption{Visual representation of the clustering. The radar plots show for a group centroid the intensity of each feature on a three classes scale. We summarize in green the topics that cover at least $85\%$ of the images categories in a cluster.} 
\label{fig:clustering}
\end{figure*}

\paragraph{Results}
We summarize in Fig. \ref{fig:clustering} the characteristics of the centroids of the $12$ most populated clusters, where each facet represents the mean value of that feature across the examples in that cluster. For ease of visualization, we discretize continuous variables in three classes: \textit{low}, \textit{medium}, or \textit{high}, according to whether the value falls, respectively, in the first, second, or third quantile of the feature distribution. To provide a more clear visual representation of the clusters, we labeled them with descriptive names. We also manually inspected the images in each cluster and chose two to four representative images among the most popular ones. A complete summary of the clustering results can be found in the Supplementary Material (Supplementary Figure).

In the rest of this section, we explore more in depth image quality and its interplay with images containing faces. Even though quality appears, on aggregate, to be moderately positively associated with the tendency to click on images, the underlying phenomenology is more nuanced. On one hand, high-quality images within the geography, transportation, visual arts, military, and biology categories (clusters 2, 3, 5, 6, 7, 8, and 9) show high iCTR across a wide range of contextual factors. A large portion of these images depicts outdoor sceneries that is coherent with the positive coefficient of the \textit{outdoor} feature in the regression in Sec.~\ref{sec:log_reg}. On the other hand, low quality images are often associated with the presence of faces, especially in topics such as biography, entertainment, and sports, wich overall tend to have a lower click-through rate. 
Focusing on the interplay between biographies and iCTR reveals significant differences across page popularity and topics worth studying. 
Images within unpopular biographies, predominantly inline and with a curated textual description, show high iCTR (cluster 10), as well as images placed in galleries in biographies of unpopular artists (cluster 1).  
On the contrary, popular biographies (cluster 11) or pages that present popular athletes (cluster 12), experience a low iCTR. A possible explanation for this behavior is that users may tend to click on an image in a biography if they do not recognize immediately the subject depicted, while for prominent celebrities, especially if the image is accessible in the infobox, the information need is fulfilled without the need of a click and the interaction with the Media Viewer.

\subsection{Are faces engaging on Wikipedia?}
\label{sec:faces}

As pointed out in Sec. \ref{sec:image_features}, images with faces generally elicit high social engagement. In Sec. \ref{sec:log_reg}, we found that the number of faces has negative weight with respect to the iCTR, while in Sec. \ref{sec:clustering} we observed that Wikipedia readers are more likely to click on images with faces only when placed in less popular biographies. To further investigate this aspect, we design a matched observational study in which we compare the iCTR between images with and without faces. To reduce the effects of confounding factors, we perform a pairwise comparison of images with similar covariates using propensity score matching.

\paragraph{Propensity score matching} 
Propensity score matching~\cite{abadie2006large} is a statistical technique to evaluate the efficacy of a treatment against a control group, while taking into account the effect of confounding factors. The propensity score is defined as the probability of a sample being treated as a function of the covariates, and it is obtained by training a logistic regression with the covariates as predictors, and the treatment/control variable as target. As a result, observations with the same propensity scores have the same distribution across the observed covariates.

In our experiment, we define images with at least one face as receiving the treatment, images without a face as the control group, and the variables used in the logistic regression (except for the topics and the page popularity) as the covariates. 

\begin{figure*}[t]
  \includegraphics[width=0.9\linewidth]{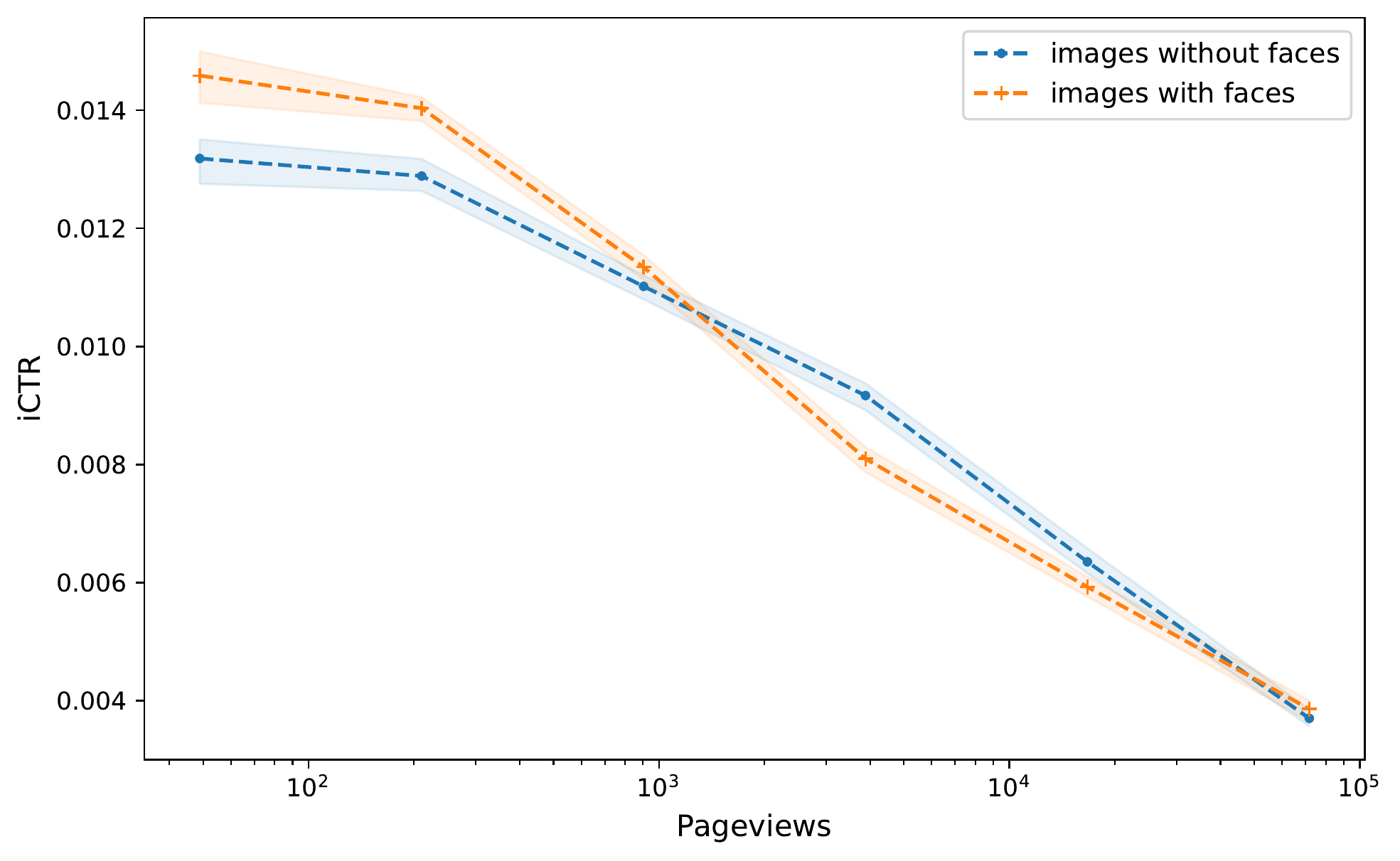}
  \caption{Comparison of the iCTR for images with faces (orange) and without faces (blue) as function of the popularity (\textit{pageviews}). Error bands represent bootstrapped $95\%$ CIs.}
  \label{fig:ictr_faces}
\end{figure*}

\paragraph{Results} We consider images in articles about biography, entertainment, and sports, accounting for $90\%$ of all images with at least one face. We find pairs of images  minimizing the propensity score within pairs of articles. Fig. \ref{fig:ictr_faces} shows the iCTR as a function of the page popularity, for images with (in orange) and without (in blue) faces. According to a Mann-Whitney U test, the difference between the two distributions is statistically significant, with $p<0.001$. The tendency to click on images with faces varies depending on page popularity. On pages with less that 1,000 monthly pageviews, the presence of faces induces higher level of interactions, with a difference of $0.1\%$, whereas, after 1,000 pageviews, we observe the opposite behavior, with a difference of $0.06\%$. This also confirms the findings of the clustering analysis.

To ascertain that our findings remain valid also for non-biographical articles, we replicate the same study by including all the topics in the matching procedure. In this case, we observe a different behavior. Images with faces are less likely to be clicked than others, across all the popularity range. This may explain the overall negative coefficient of the faces feature in the regression analysis, and highlight the role that faces play in increasing engagement on biographical articles.

%%%%% RQ3
\section{RQ3: Do images support reader's need for additional information when navigating Wikipedia?}
\label{sec:rq3}

We found that readers show a signal of interest in images when reading Wikipedia articles. But are images useful to fulfill part of the reader's information need when navigating the website? To address this question, we design an additional study that attempts to estimate whether the presence of an image in an article preview can complement the textual information and support in-depth reading.

\paragraph{Matching articles.} To check the difference in terms of conversion rate between articles having and not having an image, we first need to reduce the impact of exogenous factors that may potentially drive reader attention on articles, other than the presence of an image. For example, events localized in time can have the effect of sporadically increasing the interest towards specific articles, and therefore on the number of edits \cite{georgescu2013extracting}. Similarly, the probability of clicking on an article may also depend on its centrality in the article network, i.e. on its \textit{in-degree}, which is the number of page links pointing to that article. Ideally, we would like to find pairs of articles---one with, the other without image in the preview---that are similar in such factors. To control for these factors, we resort again to propensity score matching.
In this experiment, articles with an image in the preview are the treatment group, articles without images are the control, and we use text length, number of edits, and in degree as variables for the matching procedure. 

\begin{figure*}[t]
  \includegraphics[width=0.9\linewidth]{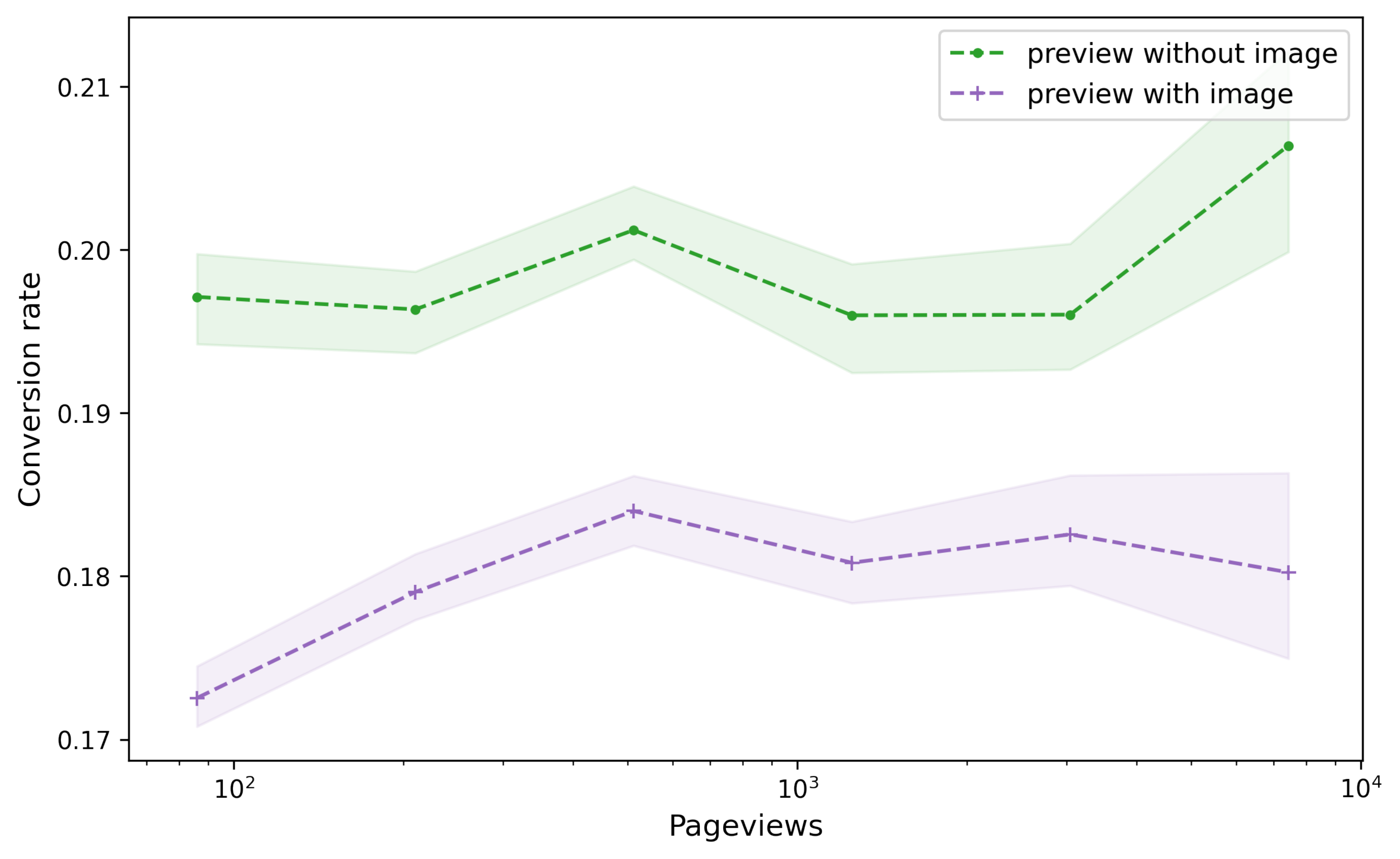}
  \caption{Comparison of the conversion rate for preview tooltip with an image (purple) and without image (green) as function of the page popularity (\textit{pageviews}). Error bands represent bootstrapped $95\%$ CIs.}
  \label{fig:conv_rate}
\end{figure*}

\paragraph{Results}
We find pairs of articles by minimizing the  propensity score within pairs of articles. 
Fig. \ref{fig:conv_rate} shows the conversion rate as a function of article popularity (total number of page views), for articles with (in blue) and without (in yellow) an image in the preview. We find that, according to a Mann-Whitney U test, the difference is statistically significant ($p<0.001$), across all the popularity spectrum, with a difference of $2\%$ in the conversion rate.

We rank all pages by conversion rate, and manually inspect the top and bottom articles, with and without images.
We find that most of the illustrated articles with higher conversion rate tend to be long lists of aggregated pieces of content related to the same topic, e,g., achievements / publications (movies, books, articles) from notable people or shows. Highly clicked illustrated page previews are often also historical events, or elections, namely information-dense articles where the lead image is only partially useful to grasp the entire article content and its complexity. Conversely, illustrated pages with low conversion rate are articles talking about a specific place (e.g., ``Old Fortress, Corfu''), or a specific person, object or spieces (e.g., ``Microvelia Macgregori''), namely articles where an illustration can satisfy most of the information need.

Unillustrated page previews with high conversion rate are much more diverse, they go from individual objects or people, e.g. (``Fanny Sidney''), where more textual information is needed to understand the subject in absence an image, to lists and events. Unillustrated articles with lower conversion rate instead tend to be about subjects where a visual explanation is not necessarily needed in order to fully understand the information: for example, generic concept such as ``Authority''. ``Miniseries'', or ``Bachelor of Science'', where images could actually be misleading or give a biased perception of the abstract piece of knowledge.

%%%%% Discussion
\section{Discussion and conclusions}
\label{sec:conclusion}
We provided a comprehensive overview over Wikipedia's visual world and how readers interact with it. We analyzed reader interactions with visual encyclopedic knowledge and found that images attract more attention than other interactive parts of the article: on average, click-through rate on images is $3.5\%$, while, for example, reference clicks happen only for $1$ in $300$ pageviews~\cite{piccardi2020quantifying}. Our insights can be summarized as follows:

\begin{itemize}
\item{\textit{Images serve a cognitive purpose.}} We found a  negative relation between article length and iCTR. This suggests that, similar to references~\cite{piccardi2020quantifying}, images might be used by readers to complement missing information in the article, fulfilling part of their \textit{cognitive} function of providing knowledge complementary to the text~\cite{levie1982effects}. 
Through a matched observational study, we also found that readers tend to click more often on unillustrated Wikipedia page previews to expand their content. On the contrary, conversion rate on illustrated page previews is consistently much lower across popularity buckets, thus suggesting that readers'  need for contextual information is often fulfilled by the presence of an image on the preview popup.
In this work, we also tested the relation betwen readers' interactions with images and article readability: our hypothesis was that images provide a \textit{compensatory} function for articles that are difficult to read. However, we found evidence of the opposite trend: more readable articles tend to elicit higher engagement with images. While this is a preliminary result, further investigation is needed to understand how images support learning in low readability contexts.
\item{\textit{We engage more with images illustrating the world and complex objects.}} Our different layers of analysis consistently expose that Wikipedia readers are attracted by images about geographic locations, especially monuments and maps, and illustrations about biological sciences. Moreover, while we did not explicitly encode the notion of image \textit{complexity} into our models, we found that Wikipedia readers tend to interact more often with images of complex objects, such as the ones in articles about visual arts, transportation, and military topics. A similar relation between the complexity of the image and its visual interestingness, i.e., the extent to which an image catches the viewer attention, has been widely explored and verified in experimental psychology and computer vision literature ~\cite{constantin2019computational}. While this relation can be influenced by different visual factors, such as the image size and its content, our results seem to support similar hypothesis, and provide a starting point for further investigation on the relation between image complexity and reader engagement.
\item{\textit{Faces engage us, but only if unfamiliar.}} Consistently, research works from different fields suggest that people and web users engage more with faces~\cite{morton1991conspec} and face pictures~\cite{bakhshi2014faces}, especially celebrities~\cite{tsikrika2014multi}, than with other objects or subject, both in online platforms and in the real world. In this work, we found an opposite trend: for Wikipedia readers, images with faces seem to be much less engaging than, for example, more ``encyclopedic'' images about monuments or transportation. However, we also found that readers do interact with face images when they are placed in unpopular articles, i.e. when those faces represent less well-known people or are \textit{unfamiliar}. This positive relation between unfamiliarity and engagement again confirms findings from previous research linking the interestingness of a visual object with its familiarity to the observer ~\cite{constantin2019computational}.
\end{itemize}
\paragraph{Implications} 
This paper represents a first step towards understanding the importance of images for free knowledge ecosystems. Inspired by theories and ideas from experimental psychology and cognitive science, and by previous studies on Wikipedia readers and web users, our findings describe for the first time how web users interact with the largest source of visual encyclopedic knowledge on the Web. These insights have several implications for different audiences.

For researchers, our results show the feasibility of large scale studies to understand the role of images in instructional settings using a multimodal, computational approach.  To this end, experiments could be designed along the same lines of this research, analyzing data coming from, for example, online learning platforms or MOOCs. Researchers could expand the depth and breath of modalities to better understand how and where images should be placed to maximize engagement and learning on the platforms.
Researchers could use this work as the basis to build predictive models for image engagement, on Wikipedia and beyond. While our work used basic visual features to understand how readers interact with images, more advanced vision techniques could be used to build end-to-end classifiers that predict the interestingness of an image for Wikipedia readers. 
This work represents a first step towards understanding the role of images in online instructional settings. While explaining the importance of images in learning is outside the scope of this work, our study shed light on how readers interact with images on Wikipedia, what attracts their attention and which types of visual content they engage with. We look at readers’ interest and usage of images using a fairly implicit, large-scale signal, namely image click-through rate. Future work looking at understanding how readers learn through Wikipedia will need to employ a different set of techniques and signals, i.e., large-scale user studies, focus groups and reading comprehension surveys. This work can be used as a starting point for learning studies. Our feature design is heavily inspired by theories from experimental psychology, computational aesthetics and educational technology research, as well as previous studies analyzing the behavior of Wikipedia readers. Researchers interested in working on learning aspects related to Wikipedia  will be able to tap into the same corpus of literature, and look into similar feature design choices.

For editors, given the large amount of unillustrated articles on Wikipedia, and the high level of interest in visual encyclopedic content, the analysis in this paper can help  editors prioritize the inclusion of visual content in areas that are highly engaging for Wikipedia readers. Longer term, models and products incoporating signals of readers' interest in visual content would be extremely helpful for editors. Tools designed to automatically predict reader engagement with images could be incorporated in services and models that help find and prioritize the right images for Wikipedia articles. Given the limited amount of information editors have about how readers interact and learn with Wikipedia content, having visibility over the potential usefuleness of an image in an article would be tremendously helpful to improve editor workflows.

For the broader Wikimedia community, the fact that images help arise interest in free knowledge justifies investments and initiatives designed to improve the pictorial representations of Wikipedia. Our findings on readers interacting with images of monuments and science further encourage the flourishing of initiatives such as Wiki Loves Monuments and Wiki Loves Science which aim at increasing the pictorial representations of these topics. Similarly, the fact that readers are more interested in pictures of unfamiliar people further justifies the existence of organizations such as ``Whose Knowledge?''\footnote{Whose Knowledge?. \url{https://whoseknowledge.org/}. Accessed March 2021.}, who pushes towards the inclusion of visual content in biographies of people from under-represented communities.

For content creators interested in contributing to free knowledge communities and in making their content available in the open, our results provides an initial list of areas of content where closing the visual knowledge gap on Wikipedia \cite{redi2020taxonomy} is crucial. Knowing that readers tend to be attracted by specific subjects and topics can help the design of new content creation campaigns and donations. The Wikimedia communities, the Wikimedia Foundation, and any web user interested in free knowledge can use these findings to collaborate with GLAM institutions and content creators to make relevant visual content free to use.

Finally, while the scope of this paper is limited to the encyclopedia, Wikipedia represents a central hub of the web ecosystem and the public domain . Its open visual content is re-used across multiple platforms and users, and its images are surfaced at the top of both text and image search results.  With this paper, we hope to provide a novel set of results and insights that can build towards better, more open and accessible visual knowledge on Wikipedia, and in turn influence the global accessibility of open visual content in the broader web.

\paragraph{Limitations} While the final goal of our research is to understand images on the broad free knowledge ecosystem, one main limitation of this work is that it mainly focuses on \textit{English} Wikipedia. With this in mind, we hope in the future to extend this work to include a more representative set of Wikipedia language editions and compare how different language communities interact with visual content.

Most of our analysis depends on the output of existing machine learning models, such as ORES~\cite{halfaker2020ores}, MTCNN~\cite{wen2016discriminative} or the novel Wikimeda Image Quality classifier. While pretty effective for this task, not all these models have been tested for fairness and inclusivity. As part of our improvements to this work, we would like to employ models that are as debiased as possible and that can be easily applied to images and articles from all around the world.

Readers from different parts of the world come to Wikipedia with different information needs~\cite{lemmerich_why_2019}. Additionally, researchers in multimedia computing have shown that different language communities \cite{pappas2016multilingual} and geographies \cite{you2017cultural} perceive and produce visual content in different ways. While we focused here on the context and content of Wikipedia images, our analysis completely ignores the characteristics of \textit{readers}, such as geographic location, internet connection availability for image download, and native language. 
Our early experiments on global reader behavior show that the way in which readers interact with images on Wikipedia tend to differ across geographic locations, mainly due to broadband availability, modality of access (mobile vs desktop), and availability of content in their languages. Previous research has indeed shown that the scope and uniqueness of visual material, as well as the availability of content for specific topics  largely varies across different language editions~\cite{he2018the_tower_of_babel,piccardi2021crosslingual}.
Our next research will extend this analysis to understand, in a privacy-preserving manner, the behavior of different groups of readers with visual encyclopedic content and the impact of exogenous events on image viewership. 

Finally, this analysis merely \textit{quantifies} reader interactions with images, without understanding the actual reason behind the action of clicking on visual content. Our choice of metrics for interest operationalization was driven by an extensive literature studying user interactions with content on web platforms, as reported in Related Work. Click-through rate and conversion rate are widely used to measure image relevance, search satisfaction, user interest in illustrated ads, and reader interactions with citations on Wikipedia. While providing a big picture of readers' behavior with Wikipedia visual content, a more detailed representation of user interactions could provide complementary insights on this front. Future work will explore a larger set of metrics such as hovers, dwell-time, and eye tracking movements. These metrics are not currently collected by the Wikipedia instrumentation pipeline and we will need to research additional data collection tools. As part of our efforts to understand the importance of images in free knowledge ecosystems, in the future we will also use surveys and user studies to learn \textit{why} readers look at images on Wikipedia, and further characterize how people use the largest visual encyclopedic knowledge repository.

%%%%%%%%%%%%%%%%%%%%%%%%%%%%%%%%%%%%%%%%%%%%%%
%%                                          %%
%% Backmatter begins here                   %%
%%                                          %%
%%%%%%%%%%%%%%%%%%%%%%%%%%%%%%%%%%%%%%%%%%%%%%

\begin{backmatter}

\section*{Supplementary material}
The PDF file entitled ``Supplementary Material'' contains:
\begin{itemize}
    \item a Supplementary Figure for the clustering analysis, depicting all the clusters not in Figure \ref{fig:clustering};
    \item a Supplementary Table including the numerical values discussed in Sec. 4.3 and 6.1.
\end{itemize}

\section*{Availability of data and materials}
The user web logs collected during the current study and the quantities computed from them are not publicly available due to privacy restrictions. All the other image features mentioned in the Data Collection section are publicly availailable at the specified urls, or are available from the corresponding author on reasonable request.

\section*{Competing interests}
The authors declare that they have no competing interests.

\section*{Funding}
RS has been partially supported by the project ``Countering Online hate speech through Effective on-line Monitoring'' funded by the Compagnia di San Paolo. The funder had no role in the study design, data collection and analysis, decision to publish, or preparation of the manuscript.

\section*{Author's contributions}
DR, TP, MR, and RS conceptualized the problem and designed research; DR, TP and MR collected data; DR ran the analysis and DR, TP, MR, and RS interpreted the results; DR, TP, MR, and RS wrote the paper. All authors read and approved the final manuscript.

\section*{Acknowledgements}
The authors thank the Wikimedia Research Team for their insightful discussions and the Analytics Team for their technical support.

%%  After submission of the .TEX file,                     %%
%%  you will be prompted to submit your .BBL file.         %%

% if your bibliography is in bibtex format, use those commands:
\bibliographystyle{bmc-mathphys} % Style BST file (bmc-mathphys, vancouver, spbasic).
\bibliography{bibliography}      % Bibliography file (usually '*.bib' )
% for author-year bibliography (bmc-mathphys or spbasic)
% a) write to bib file (bmc-mathphys only)
% @settings{label, options="nameyear"}

% or include bibliography directly:
% \begin{thebibliography}
% \bibitem{b1}
% \end{thebibliography}

\end{backmatter}

\end{document}